\documentclass[11pt]{article}
      
\usepackage{a4wide}
\usepackage{amsmath}
\usepackage{amsfonts}
\usepackage{amsthm}
\usepackage{amssymb}
\usepackage[dvips]{graphicx}
\usepackage{exscale}
\usepackage{color}

\theoremstyle{definition}

\newcommand{\mR}{\mathbb{R}}

\newcommand{\mZ}{\mathbb{Z}}
\newcommand{\mN}{\mathbb{N}}

\newcommand{\Gp}[1]{G^+_{#1}}
\newcommand{\Gm}[1]{G^-_{#1}}

\newcommand{\tgp}[1]{\widetilde{G}^+_{#1}}
\newcommand{\tgm}[1]{\widetilde{G}^-_{#1}}

\newcommand{\sA}{\mathcal{A}}
\newcommand{\sC}{\mathcal{C}}

\newcommand{\sN}{\mathcal{N}}
\newcommand{\sS}{\mathcal{S}}

\newcommand{\sV}{\mathcal{V}}

\newcommand{\wt}[1]{\widetilde{#1}}

\newcommand{\ct}{\tilde{c}}
\newcommand{\zt}{\tilde{z}}

\newcommand{\ket}[1]{|#1\rangle}

\newcommand{\flowket}[1]{|#1\rangle_\eta}
\newcommand{\etaket}{\flowket{h^\eta,q^\eta}}

\newcommand{\vpp}{\vphantom{+}}

\DeclareMathOperator{\sgn}{sgn}
\DeclareMathOperator{\mTr}{Tr}
\newcommand{\Tr}[1]{\mTr^{\vpp}_{#1}}

\numberwithin{equation}{section}

\begin{document}

\begin{titlepage}

\flushright{KCL-MTH-03-07}

  \begin{center}
    
    \vspace{2cm}
    
    \Large{\textbf{Embedding Diagrams of the N=2 Superconformal
        Algebra under Spectral Flow}}
    
    \vspace{2cm}
    \large{ Hanno Klemm\footnote{\tt{klemm@mth.kcl.ac.uk}}

      \vspace{0.3cm}
      Department of Mathematics \\
      King's College London \\
      Strand\\
      London WC2R 2LS \\}

    \vspace{1cm}
\begin{abstract}
  \noindent
  The embedding diagrams of representations of the {$N=2$}
  superconformal algebra with central charge $c=3$ are given. Some
  non-unitary representations possess subsingular vectors that are
  systematically described. The structure of the embedding diagrams is
  largely defined by the spectral flow symmetry. As an additional
  consistency check the action of the spectral flow on the characters
  is calculated.
\end{abstract}

\end{center}
\end{titlepage}

\section{Introduction}
\label{sec:intro}

The representation theory of the $N=2$ superconformal algebra was
long thought to be an obvious generalisation of the representation
theory 
of the Virasoro and $N=1$ superconformal algebra. It was first shown
in \cite{doerrzapf,Dorrzapf:1995wv,Dorrzapf:1994es} that this is not
quite the case (see \cite{Dobrev:1987hq} for some earlier results). In
particular the    
structure of singular vectors turned out to be more involved than in
the $N=1$ supersymmetric generalisation of the Virasoro algebra. 

In \cite{Gato-Rivera:1996ma} it was noticed that some representations
of the $N=2$ superconformal algebra contain subsingular 
vectors. However, only some explicit examples are known which are all
constructed using the ``topological twisted'' algebra and then
translated back via the topological ``untwisting''. The examples found
in this way showed that subsingular vectors exist but did not
admit a systematic exploration of these. 

In \cite{Semikhatov:1998gv,Semikhatov:1997pf,Feigin:1998sw} the
representation theory of the topologically twisted $N=2$
superconformal algebra was treated 
via the representation theory of $\widehat{sl}(2)$. By this approach
some of the embedding diagrams for $c=3$ were found in
\cite{Semikhatov:1997pf} however, as the relation of the $N=2$
superconformal algebra with $\widehat{sl}(2)$ representations breaks
down at $c=3$ not all embedding diagrams could be approached in this
way. In \cite{Semikhatov:1998gv} a classification of all subsingular
vectors for $c<3$ was tried, however, as was pointed out in
\cite{Gato-Rivera:1999ey}, some cases seem to have been overlooked. 

Our main aim in this paper is to clarify the structure of the embedding
diagrams for central charge $c=3$. To this end we pursue another route.
The $N=2$ superconformal algebra possesses a family of outer
automorphisms $\alpha_\eta$, called the spectral flow, which map the
algebra to itself. For integer values of the flow parameter $\eta$,
the Neveu-Schwarz sector is mapped to itself as is 
the Ramond sector. We study the induced action of the spectral flow on
the representations of the algebra. 
It is shown that the representations are mapped onto each other under
spectral flow in a 
systematic manner. In particular the embedding diagrams can be grouped
into 
a few categories which transform among themselves under spectral flow.
Furthermore we obtain an algorithm to derive subsingular
vectors from singular vectors via spectral flow transformations. We
show that only non-unitary representations posses subsingular vectors.
Although we only cover the representations with $c=3$ in detail, we 
briefly demonstrate that our techniques are applicable to
representations with other values of the central charge in the case of the
$N=2$ unitary minimal models. 

The paper is organised as follows. In chapter \ref{sec:n2algebra} we
review some well known facts about the $N=2$ algebra and introduce the
spectral flow to fix our conventions. In chapter
\ref{sec:rep-flow} we explore how the spectral flow acts on a
given representation and how 
representations are transformed. In Chapter \ref{sec:singular-flow} 
we describe how singular vectors behave under spectral
flow and how subsingular vectors arise under spectral flow
transformations. In chapter
\ref{sec:ramond-algebra} we comment briefly on how the Ramond algebra can
be analysed by the same method.
In chapter \ref{sec:characters-c=3} we
derive the characters of a large class of representations and
determine the action of the spectral flow on them. In order to check
the consistency of the embedding diagrams derived above we also construct 
the characters for representations with subsingular vectors.  In
chapter \ref{sec:minimal} we briefly describe how the same methods can
be used for the unitary minimal models.   
Chapter \ref{sec:conclusions} contains
further comments and outlooks. Some technical proofs can be found in
the appendix.  

\section{The $N=2$ Algebra and Spectral flow}
\label{sec:n2algebra}

\subsection{Basic facts}
\label{sec:basic}

We want to review some basic facts of the $N=2$ superconformal algebra
before we proceed. 

The $N=2$ superconformal Algebra ${\cal A}$
consists of the Virasoro algebra $\{L_n\}$, a  weight one
$U(1)$-current $\{J_n\}$ and the modes of two supersymmetric partners 
$\{G_r^\pm\}$ of conformal dimension $h=\tfrac{3}{2}$, obeying the
(anti-)commutation relations \cite{Ademollo:1976an}
\begin{align}
  \left[L_m, L_n \right] &=(m-n)L_{m+n} + \tfrac{\ct}{4}(m^3 -
  m)\delta_{m,-n} \\ 
  \left[L_m, J_n \right] &=-nJ_{m+n}\\
  \left[L_m, G_r^\pm \right] &=\left( \tfrac{1}{2}m-r \right)
  G_{m+r}^\pm \\
  \left[J_m, J_n \right] &= \ct m\delta_{m,-n}\\
  \left[J_m, G_r^\pm \right] &=\pm G_{r+m}^\pm \\
  \left\{ \Gp{r}, \Gm{s}\right\} &= 2 L_{r + s} + (r -s) J_{r + s} +
  \ct(r^2 - \tfrac{1}{4})\delta_{r, -s} \\
  \left\{\Gp{r}, \Gp{s} \right\} &=\left\{\Gm{r}, \Gm{s}\right\}=0
\end{align}
where $\ct=\tfrac{c}{3}$ and braces denote anticommutators. In the NS
sector the modes $L_n$ and $J_n$ are integral and the  modes $G_r^\pm$
are half-integral whereas in the R sector all modes are integral. 

The determinant formula of the NS sector of this
algebra at level $n$ with relative charge $m$ was first written down in
\cite{kent} and proved in \cite{Kato:1987td}. It is given by
\begin{equation}
  \label{eq:determinant}
  \det M_{n,m}^A(\ct, h, q) 
  = \prod_{1\leq rs\leq 2n \atop s \text{ even}}
  (f_{r,s}^A)^{P_A(n-rs/2,m)}
  \prod_{k \in \mZ + 1/2}(g_k^A)^{\tilde{P}_A(n - |k|, m - \sgn(k);k)},
\end{equation}
where $P_A$ is defined by
\begin{equation}
  \sum_{n,m} P_A(n,m)x^ny^m = \prod_{k=1}^\infty 
\frac{( 1 +  x^{k-1/2} y)(1 + x^{k-1/2}y^{-1})}{(1-x^k)^2}.
\end{equation}
$P_A(n,m)$ counts the states at level $n$ with relative charge $m$.
$\tilde{P}_A(n,m,k)$ is given by
\begin{equation}
   \sum_{n,m} \tilde{P}_A(n,m,k)x^n y^m = (1+ x^{|k|}y^{\sgn(k)})^{-1}
   \sum_{n,m} P_A(n,m)x^ny^m,
\end{equation}
with $\sgn(k)=1$ for $k>0$ and $\sgn(k)=-1$ for
$k<0$. $\tilde{P}_A(n,m,k)$ counts the states build on 
vectors with relative charge $\sgn(k)$ at level $k$. The functions
$f^A$ and $g^A$ are given by
\begin{align}
  f_{r,s}^A(\ct, h, q) &= 2(\ct - 1)h - q^2 - \tfrac{1}{4}(\ct -1)^2 + 
  \tfrac{1}{4}((\ct -1)r + s)^2, & s \text{ even} \\
  g_k^A(\ct, h, q) &= 2h - 2kq + (\ct -1)(k^2 - \tfrac{1}{4}), & k \in
  \mZ+\tfrac{1}{2}. 
\end{align}
If we set $\ct = 1$ these formulas reduce quite drastically to 
\begin{align}
  f_{r,s}^A(\ct=1, h, q) &= \tilde{s}^2 - q^2, & \tfrac{s}{2} =
  \tilde{s} \in \mZ \\ 
  g_k^A(\ct=1, h, q) &= 2(h - kq),& k \in \mZ+\tfrac{1}{2},
\end{align}
and the determinant formula then reads
\begin{equation}
  \label{eq:determinant2}
  \det M_{n,m}^A(\ct=1, h, q) 
  = \prod_{1\leq rs\leq 2n \atop s \text{ even}} 
  (\tfrac{1}{4}s^2 - q^2)^{P_A(n-rs/2,m)}
  \prod_{k \in \mZ + 1/2}2(h - kq)^{\tilde{P}_A(n - |k|, m - \sgn(k);k)}.
\end{equation}

\subsection{Representations of $N=2$ at $c=3$}
\label{sec:reps}

We will denote \emph{Verma modules} by $\sV_{h,q}$ where $h$ and $q$
denote weight and charge of  
the highest weight vector (hwv). They are in general
not irreducible representations but
contain \emph{null vectors} i.e.\  vectors whose inner product with
any other vector vanishes. These vectors have
to be quotiened out in order to obtain an irreducible representation. 
Null vectors which by themselves are highest weight states, i.e.\  which
are annihilated by the action of any lowering operator, are called
\emph{singular vectors}. They span submodules inside a Verma module. 
The ${N=2}$ superconformal algebra furthermore possesses \emph{subsingular
vectors} which are null vectors that are neither singular vectors nor
descendants of singular vectors. Once the singular vectors are
quotiened out subsingular vectors become (new) singular vectors.

Any highest weight
representation is determined by the position of its singular
(and subsingular) vectors.
Vanishing of the determinant formula (\ref{eq:determinant2}) signals
null vectors at the given level and relative charge. At the level
where vanishings first occur we then find a singular vector. 
As was already pointed out in \cite{kent},
singular vectors can only exist with relative charge $m = 0, \pm 1$. 

The position and the relations
among the singular vectors can be summarised in the \emph{embedding
  diagram} of a given 
representation. We want to study the relations among the embedding
diagrams under spectral flow transformations. 
The embedding diagrams of various highest weight representations were
calculated in \cite{doerrzapf}. Unfortunately some of the results
are incomplete as the assumption made that the representations of the
$N=2$ superconformal algebra do not contain subsingular vectors
does not hold. The first paper where subsingular vectors of the $N=2$
superconformal algebra were discovered was \cite{Gato-Rivera:1996ma}.  
The structure of the embedding diagrams depends mainly on the
\begin{enumerate}
\item value of $q$: Only for $q\in \mZ$ (and $q \neq 0$ ) uncharged
  singular vectors exist, as can be seen from the determinant
  formula \eqref{eq:determinant2}, in particular from the form of $f^A$. 
\item value of $\tfrac{h}{q}$: Only if this value is half integer
  $g^A$ has zeros and thus only then charged singular vectors exist.
\end{enumerate}
There are a few different classes of representations we have to consider
in the case of central charge $c=3$. 

All representations with charged singular vectors appear in pairs: to
any representation with charged singular vectors there exists a
representation with the same embedding structure in which the relative
charge of the singular vectors is reversed. 
This is a
consequence of the mirror automorphism \cite{Lerche:1989uy,
  Greene:1990ud, Greene:1996cy} which connects representations that
differ only by 
the sign of the $U(1)$ charge. These representations are mapped onto
each other by the mirror automorphism $m$ acting on the modes as follows
\begin{align}
  \label{eq:19}
  m(L_n) &= L_n \\
  m(J_n) &= -J_n, \\
  m(G^\pm_r) &= G^\mp_r.
\end{align}
Therefore any statement about a representation with a given charge $q$
holds for the representation with charge $-q$, as well.  

\begin{figure}[htbp]
  \begin{center}
    \includegraphics[width=6cm]{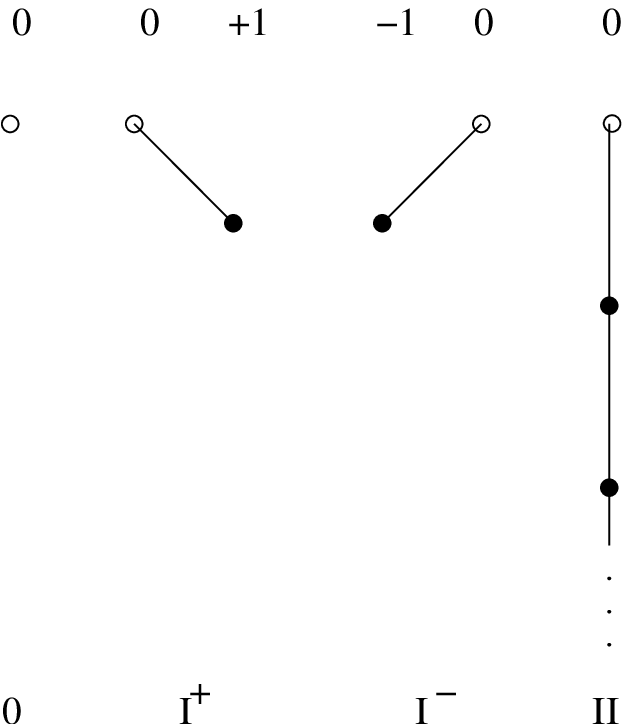}
    \caption{The embedding diagrams for the $c=3$ representations
      considered here: 
      $\text{0}$: $q \not\in\mZ$, $\tfrac{h}{q} \not\in \mZ +
      \tfrac{1}{2}$, or $q=0$, $h\neq 0$;
      I$^+$: $q \not\in\mZ$, $\tfrac{h}{q} \in \mZ^+ +
      \tfrac{1}{2}$;
      $\text{I}^-$: $q \not\in\mZ$, $\tfrac{h}{q} \in \mZ^- +
      \tfrac{1}{2}$;
      $\text{II}$: $q \in \mZ$ , $\tfrac{h}{q} \not\in \mZ + \tfrac{1}{2}$}
    \label{fig:others} 
  \end{center}
\end{figure}

\begin{figure}[htbp]
  \begin{center}
    \includegraphics[height=8cm]{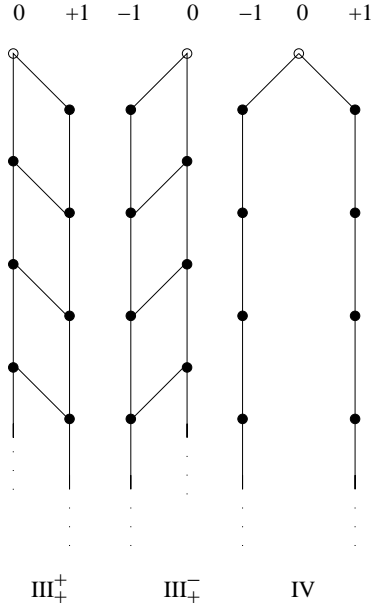}
    \caption{The embedding diagrams for $c=3$ representations
      with $\tfrac{h}{q} \in \mZ + \tfrac{1}{2}$ and 
      III$_{+}^{+}:{h,\,q>0}$;
      III$_{+}^{-}:h>0,\,q<0$; 
      IV: $h=q=0$.}   
    \label{fig:embeddingpositiv}
  \end{center}
\end{figure}

\begin{figure}[htbp]
  \begin{center}
    \includegraphics[width=13cm]{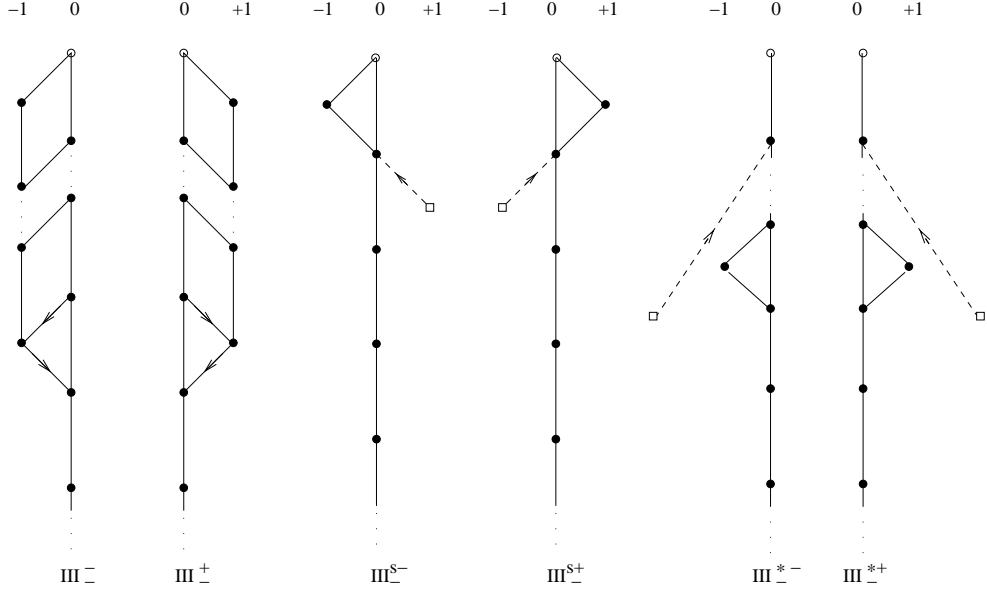}
    \caption{The embedding diagrams for $c=3$ representations
      with $\tfrac{h}{q} \in \mZ + \tfrac{1}{2}$ and 
      $\text{III}_{-}^{-}$: $h<0,\, q \geq 2$;
      $\text{III}_{-}^{+}$: $h<0,\,q \leq -2$; 
      $\text{III}_{-}^{s-}$: $h=-q/2,\, q \geq 1$;
      $\text{III}_{-}^{s+}$: $h=q/2,\,q \leq -1$;  
      $\text{III}_{-}^{*-}$: $h<0,\, q=1$;
      $\text{III}_{-}^{*+}$: $h<0,\, q=-1$.} 
    \label{fig:embeddingsub}
  \end{center}
\end{figure}

In figure \ref{fig:others} we have summarised the
embedding diagrams for which either $q \not\in \mZ$ or $\tfrac{h}{q} \not\in
\mZ + \tfrac{1}{2}$. These diagrams were already given in \cite{doerrzapf}.
For the remaining cases, i.e for the cases where $\tfrac{h}{q} \in \mZ
+ \tfrac{1}{2}$ and $q \in \mZ$ we conjecture that the form of the
embedding diagrams is as shown in figures \ref{fig:embeddingpositiv}
and \ref{fig:embeddingsub}. In an embedding diagram a black dot
denotes a singular vector and a 
line marks an operator connecting two singular vectors. The
unfilled circles denote highest weight vectors. In figure
\ref{fig:embeddingsub} the 
boxes mark subsingular vectors and the dashed lines indicate to
which singular vector the subsingular vectors are
linked\footnote{This can be ambiguous, see section \ref{sec:more-sub}.}.
In the
presence of subsingular vectors we have marked some lines with arrows
to denote which vectors are connected by operators. In all cases of
singular vectors a raising operator connects a singular vector of
lower level to a singular vector of higher level. In the case of
subsingular vectors a lowering operator maps the subsingular vector to
some (descendant of a) singular vector.

In figure \ref{fig:embeddingpositiv} and \ref{fig:embeddingsub} the
relative charge of the charged singular vectors is given by the sign
of $h/q$. Uncharged singular vectors arise at levels $n|q|$, charged
singular vectors are given at levels $\left|\tfrac{h}{q}\right| + n|q
+ \sgn(q)|$. The  
diagrams III$^{\pm}_{+}$ (figure \ref{fig:embeddingpositiv}) contain  
infinitely many uncharged and charged singular vectors. 
Diagram
IV in figure \ref{fig:embeddingpositiv} is the embedding diagram for
the vacuum representation. It contains two charged singular vectors at
each level of the form $\ell = \tfrac{1}{2} + k, \, k \in \mN$ with
relative charge $\pm1$. The embedding
diagrams given in figure \ref{fig:embeddingsub} are the embedding
diagrams of the representations with $h<0, \tfrac{h}{q}\in
\mZ+\tfrac{1}{2}, q\in\mZ$. 

Embedding diagrams of this kind contain only finitely many charged
singular vectors. The representations with $h<0, q\in \mZ$ and
$|\tfrac{h}{q}|=\tfrac{1}{2}$ contain subsingular vectors. We have
denoted these embedding diagrams by III$_-^{s\pm}$. Representations
with $h=-\tfrac{2l + 1}{2}, l \in \mN$ and $|q|=1$ constitute another
special case. They have only one charged singular vector and possess
subsingular vectors of higher charge. The associated embedding
diagrams are denoted by III$_-^{*\pm}$. We have adopted the notation that the
superscript on the label of a diagram denotes the relative charge of
the charged 
singular vectors and a subscript denotes whether the value of $h$ is
greater or smaller than zero in cases where the structure of the
embedding diagrams is affected by the sign of $h$.

Note that subsingular vectors arise only for representations with
$h<0$ and that there exist
subsingular vectors with relative charge greater than one. We were not able to
prove the actual form of the embedding diagrams but we will show that
they are compatible with the spectral flow. Explicit calculations
further support our conjectures. 
For the various values of $h$ and $q$ we obtain the following
diagrams:

\bigskip
\noindent
\begin{tabular}{l c cl}
 & \emph{levels of uncharged} & \emph{levels of charged} & \\
\emph{$h$ and $q$} & 
\emph{singular vectors} &
\emph{singular vectors} & 
\emph{diagram} \\ 
&&\\
$q \not\in\mZ$, $\tfrac{h}{q} \not\in \mZ + \tfrac{1}{2}$ 
& none & none &  0 \\
$q \not\in\mZ$, $\tfrac{h}{q} \in \mZ + \tfrac{1}{2}$ 
& none & $\left|\tfrac{h}{q}\right|$ 
&  $\text{I}^\pm$ \\
$q \in \mZ$, $q\neq 0$ , $\tfrac{h}{q} \not\in \mZ + \tfrac{1}{2}$ & 
$n|q|, n\in \mN$ & none &  II \\
$q=0, \quad h\neq 0$ & none & none &  0 \\

$q\in \mZ^+$, $h>0$, $\tfrac{h}{q} \in \mZ^+ + \tfrac{1}{2}$
& $nq, n\in \mN$ 
& $\tfrac{h}{q} + k(q+1), \,k \in \mN^0$ 
& III$^+_+$ \\
$q\in \mZ^-$, $\tfrac{h}{q} \in \mZ + \tfrac{1}{2}$, $h>0$ 
& $n|q|, n\in \mN$ 
& $\left|\tfrac{h}{q}\right| + k(|q|+1), \,k \in \mN^0$ 
& III$^-_+$ \\ 
$q=h=0$ & none & $k + \tfrac{1}{2} \,k \in \mN^0$ 
& IV \\
$q\in \mZ^+$, $\tfrac{h}{q} \in \mZ + \tfrac{1}{2}$, $h<0$ & $nq, n\in \mN$ 
& $\ell:=\left|\tfrac{h}{q}\right| + k(q-1), \ell\leq |h|$ 
&  III$^{-}_{-}$, III$^{s-}_{-}$ \\
$q\in \mZ^-$, $\tfrac{h}{q} \in \mZ + \tfrac{1}{2}$, $h<0$ 
& $n|q|, n\in \mN$ 
& $\ell:=\tfrac{h}{q} + k(|q|-1), \ell\leq |h|$ &  III$^{+}_{-}$,
III$^{s+}_{-}$ \\
$q\in \mZ^+$, $\tfrac{h}{q} =-\tfrac{1}{2}$, $h<0$ & $nq, n\in \mN$ 
& $\ell:=\left|\tfrac{h}{q}\right| + k(q-1), \ell\leq |h|$ 
&  III$^{s-}_{-}$ \\
$q\in \mZ^-$, $\tfrac{h}{q} = \tfrac{1}{2}$, $h<0$ 
& $n|q|, n\in \mN$ 
& $\ell:=\tfrac{h}{q} + k(|q|-1), \ell\leq |h|$ &  III$^{s+}_{-}$ \\
$q=1$, $\tfrac{h}{q} \in \mZ + \tfrac{1}{2}$, $h<0$ & $nq, n\in \mN$ &
$\left|\tfrac{h}{q}\right|$ &  III$^{*-}_{-}$ \\
$q=-1$, $\tfrac{h}{q} \in \mZ + \tfrac{1}{2}$, $h<0$ & $n|q|, n\in \mN$ & $\tfrac{h}{q}$ &  III$^{*+}_{-}$
\end{tabular}

\subsection{Spectral Flow}
\label{sec:flow}

There exists a family of outer automorphisms $\alpha_\eta: \sA
\rightarrow \sA$ which map the $N=2$ superconformal algebra to
itself. This family of automorphisms is called {\em spectral
  flow} \cite{schwimmer, Lerche:1989uy}. Its action on the modes is given by 
\begin{align}
  \alpha_\eta(\Gp{r}) & = \tgp{r} = \Gp{r-\eta} \\
  \alpha_\eta(\Gm{r}) & = \tgm{r} = \Gm{r+\eta} \\
  \alpha_\eta(L_n) & = \wt{L}_n = L_n - \eta J_n +
  \ct\tfrac{\eta^2}{2} \delta_{n,0}\\ 
  \alpha_\eta(J_n) & = \wt{J}_n = J_n -  \ct \eta \delta_{n,0},
\end{align}
where $\eta \in \mR$ is called \emph{flow parameter}. $\eta \in
\mZ$ maps the NS 
and R sectors  of the algebra to themselves whereas for $\eta \in \mZ +
\tfrac{1}{2}$ the spectral flow maps the NS sector to the R sector and
vice-versa. We will almost exclusively study the case in which $\eta
\in \mZ$ acts on the NS sector. Therefore from now on $\eta$ is
assumed to be an integer unless otherwise stated.

The action of the spectral flow on the operators of the algebra
induces an action on the representations. We will take the
point of view that under a spectral flow transformation the highest
weight state we build representations upon is unchanged. However the
action of the transformed modes on the same state gives rise to a new
representation. In particular with respect to the new modes the
previous highest weight vector might become a descendant state.
This point of view has the advantage that the inner product of the vector
space the modes act upon is manifestly unchanged. This implies that
null vectors must be mapped to null vectors under spectral flow.
As an example let us consider the spectral flow with
flow parameter $\eta = 1$ acting on some representation $(h,q)$. By
fixing $\ct=1$ we obtain 
\begin{align}
  \label{example}
  \alpha_1(\Gp{1/2})\ket{h,q}&=\tgp{1/2}\ket{h,q}=
  \Gp{-1/2}\ket{h,q} \\
   \alpha_1(\Gm{1/2})\ket{h,q}&=\tgm{1/2}\ket{h,q}=
  \Gm{3/2}\ket{h,q} \label{example2} \\
  \alpha_1(L_0)\ket{h,q} &=\wt{L}_0\ket{h,q}=(L_0 - J_0 +
  \tfrac{1}{2} )\ket{h,q} = (h - q + \tfrac{1}{2})\ket{h,q} \\
  \alpha_1(J_0)\ket{h,q}& =\wt{J}_0\ket{h,q}=(J_0 - 1)\ket{h,q} =
  (q-1)\ket{h,q}.
\end{align}
This example already reveals a crucial observation, namely the
spectral flow does not respect the highest weight condition because in
general $\alpha_1(\Gp{1/2})\ket{h,q}\neq 0$.

Let now be $\eta \in \mN$ arbitrary. The highest
weight condition for the fermionic operators expressed in the
transformed modes is given by 
\begin{equation}
  \label{eq:hwc}
  \forall r>0: \tgm{r}\ket{h,q} = \tgp{r}\ket{h,q} =0.
\end{equation}
If we express the transformed modes in terms of the original modes 
condition (\ref{eq:hwc}) becomes
\begin{align}
  \forall r > 0:\, \Gm{r+\eta} \ket{h,q} & = 0 \\
  \forall r > 0:\, \Gp{r-\eta} \ket{h,q} & = 0 \label{eq:8}
\end{align}
from which the second condition is not satisfied for $\eta \in
\mN$. Therefore the hwv 
changes under spectral flow and a descendant in the original module
serves as the highest weight vector of the transformed modes. The
highest weight state for the transformed modes is generically given by
\begin{equation}
\label{eq:newhwv}
  \Gp{-\eta + 1/2} \dots \Gp{-1/2}\ket{h,q} =: \etaket, \qquad \eta\in
  \mN.
\end{equation}
For $\eta \in \mZ, \eta < 0$ the $\Gp{s}$ in formula (\ref{eq:newhwv})
have to be interchanged with $\Gm{s}$. 
With respect to the original modes the weight and charge of $\etaket$
are given by 
\begin{align}
  L_0\etaket & = (h + \tfrac{\eta^2}{2} )\etaket \\
  J_0\etaket & = (q + \eta)\etaket. 
\end{align}
The weight and charge of $\etaket$ with respect to the transformed
modes are given by
\begin{align}
   \wt{L}_0\etaket
  & = (L_0 - \eta J_0 + \tfrac{\eta^2}{2})\etaket = (h - \eta q)
  \etaket
  \label{eq:newweight} \\
  \wt{J}_0 \etaket & = q \etaket. \label{eq:newcharge}
\end{align}
Let us for later convenience define
\begin{align}
  \wt{L}_0\etaket &= h^\eta \etaket,  & h^\eta &= h - \eta q  \quad
  \text{and} \label{eq:23} \\
  \wt{J}_0\etaket &= q^\eta \etaket,   &q^\eta &=
  q \label{eq:24}. 
\end{align}
The charge of the highest weight state generically does  not
change under spectral flow. Note that we defined $h^\eta$ and $q^\eta$
to denote the eigenvalues of $\wt{L}_0$ and $\wt{J}_0$
respectively. 

There exists one exception to this construction
namely if equation \eqref{eq:8} is satisfied in the sense
that the state defined by equation \eqref{eq:newhwv} is a null state.
This was discussed in a slightly different context for $c<3$ in
\cite{Gaberdiel:1997kf}. 
In this case the ``correction'' of the highest weight state is not
necessary and for the case $\eta=1$ the new highest weight state has
weight and charge given by 
\begin{align}
  \label{eq:9}
  \wt{L}_0 \etaket & = (L_0 - J_0 + \tfrac{1}{2})\ket{h,q}
  =(h-q+\tfrac{1}{2})\ket{h,q}, \\ 
  \wt{J}_0 \etaket & = (J_0 - 1)\ket{h,q} = (q-1)\ket{h,q}. \label{eq:10}
\end{align}
The formula for general $\eta$ is slightly more complicated.

Unless $G^\pm_{-1/2}\ket{h,q}$ is a singular vector, 
$\ket{h^\eta, q^\eta}_\eta$
can never be a descendant of a singular vector in the original
representation. This can be seen as follows. 
Let $n$ denote the level and $m$ the charge of a vector. For
any given level $n$ there is a maximal value for the ratio
$\tfrac{m}{n}$. Vectors of the form \eqref{eq:newhwv} are exactly
the vectors for which the ratio $\tfrac{m}{n}$ is maximal at
their level. On the other hand singular vectors have at most charge
one at some level $n_s$. Therefore any descendant null vector has
a lower ratio of $\tfrac{m}{n}$.
The only singular vectors whose descendants could have a maximal ratio
of  $\tfrac{m}{n}$ are exactly
$G^\pm_{-1/2}\ket{h,q}$. Therefore if $G^\pm_{-1/2}\ket{h,q}$ is not
singular, $\etaket$ will not be a singular vector
with respect to the original representation.

The construction of new singular vectors under spectral flow will prove to
be essentially the same as for the highest weight state. As we shall
see, the situations where 
the highest weight state is not shifted under spectral flow sometimes gives
rise to subsingular vectors. This happens in particular in the
non-unitarizable representations of type III.

\section{Representations under spectral flow}
\label{sec:rep-flow}

We now discuss how the spectral flow acts on representations. As an
easy exercise we will first address the action on the embedding
diagrams shown in figure \ref{fig:others} before we turn to the
representations shown in the other diagrams. Our aim is to identify
which representations are connected under spectral flow and therefore
to group the various representations into orbits under spectral flow
transformations. These orbits will depend on the charge $q$ of the
highest weight states and the relative sign between the highest weight
and the charge.

\subsection{The easy case}
\label{sec:easy-case}

Let us first turn to diagrams without any singular vectors, that is to
diagrams of type $0$ in figure \ref{fig:others}. The spectral flow
only shifts the highest weight vector to another vector in the Verma
module. The construction of the highest weight vector is as shown in
equation 
\eqref{eq:newhwv} and the connection between the representations is the
following
\begin{equation}
  \label{eq:18}
  \ket{h^\eta,q^\eta}_\eta = \ket{h - \eta q,q}
\end{equation}
for all values of $\eta$. 

The situation is analogous for diagrams of type II\@. The charge $q$
of the highest weight state does not change and therefore the position
of the singular vector remains unchanged and equation \eqref{eq:18}
describes the orbits of these representations, as well.

For diagrams of type I the situation is slightly more
involved. As long as $\left|\tfrac{h}{q}\right|\neq\tfrac{1}{2}$ the
situation is unchanged. However for
$\left|\tfrac{h}{q}\right|=\tfrac{1}{2}$  one of the states
$G^\pm_{-1/2}\ket{h,q}$ is a singular vector. For definiteness let us
choose $\Gp{-1/2}\ket{h,q}$ to be the singular vector. If we now
perform a spectral flow transformation with $\eta=1$ the highest
weight condition is satisfied up to the singular vector. Therefore the
highest weight state after spectral flow is given by the original
highest weight state and has highest weight and charge according to
equations \eqref{eq:9} and \eqref{eq:10}. 
Each pair of values for $h$ and $q$ such that
$\left|\tfrac{h}{q}\right|=\tfrac{1}{2}$ gives rise to an orbit of the
spectral flow. Under
spectral flow these representations are then mapped to representations
whose value of $h$ differs by integer values
\begin{equation}
  \label{eq:22}
    \ket{\tfrac{q}{2}^\eta,\, q^\eta}_{\eta} = \begin{cases}
    \ket{-\tfrac{q-1}{2} - (\eta -1)(q-1),\, q-1}, & \text{for }
    \eta > 0 \\ 
    \ket{\tfrac{q}{2} - \eta q,\, q}, & \text{for } \eta < 0 .
  \end{cases}
\end{equation}

\subsection{The interesting case}
\label{sec:interesting-case}

Let us now specify to values of $h$ and $q$ such that $q \in \mZ$ and
$\tfrac{h}{q}\in \mZ + \frac{1}{2}$. These representations correspond
to the embedding diagrams of type III of figures
\ref{fig:embeddingpositiv} and \ref{fig:embeddingsub}. The
representations corresponding to diagrams of this type have 
uncharged singular vectors at level $n|q|$ with $n\in \mN$ and
charged singular vectors of relative charge $\text{sgn}(\tfrac{h}{q})$
of which the first is at level 
$\left|\tfrac{h}{q}\right|$. 
Spectral flow of these representations gives as new value for
$\tfrac{h^\eta}{q^\eta}$ 
\begin{equation}
  \frac{h - \eta q}{q} = \frac{h}{q} - \eta \in \mZ + \frac{1}{2}.
\end{equation}
Therefore representations of this class are mapped onto each other
under spectral flow. In particular as long as the value of $h^\eta$
does not change sign with respect to $h$, the value of $q$ remains
unchanged.

However, the existence of singular vectors  might spoil the
construction of the highest weight vector in some cases as described
in section \ref{sec:flow}. For $\eta=1$ the weight
and charge  
with respect to the transformed modes are then given by equations
\eqref{eq:9} and \eqref{eq:10} such that 
\begin{equation}
  \frac{\alpha_1(L_0)\ket{h,q}}{\alpha_1(J_0)\ket{h,q}}=\frac{h^\eta}{q^\eta}
  = -\frac{1}{2}.
\end{equation}
This indicates that the transformed representation has a
negatively charged vector at level $\tfrac{1}{2}$. This was to be expected
as this singular vector is needed to reach the original representation
by the inverse spectral flow transformation.

\begin{figure}[htbp]
  \centering
  \includegraphics[width=12cm]{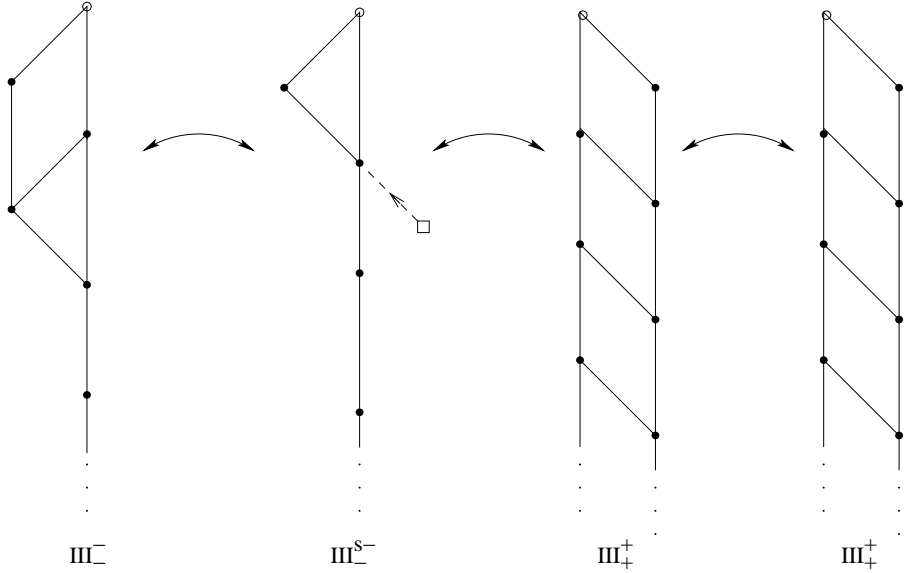}
  \caption{Spectral flow of type III diagrams, for $h>0, |q|>2$ to
  $h<0$}
  \label{fig:typeIII}
\end{figure}

\begin{figure}[htbp]
  \centering
  \includegraphics[width=12cm]{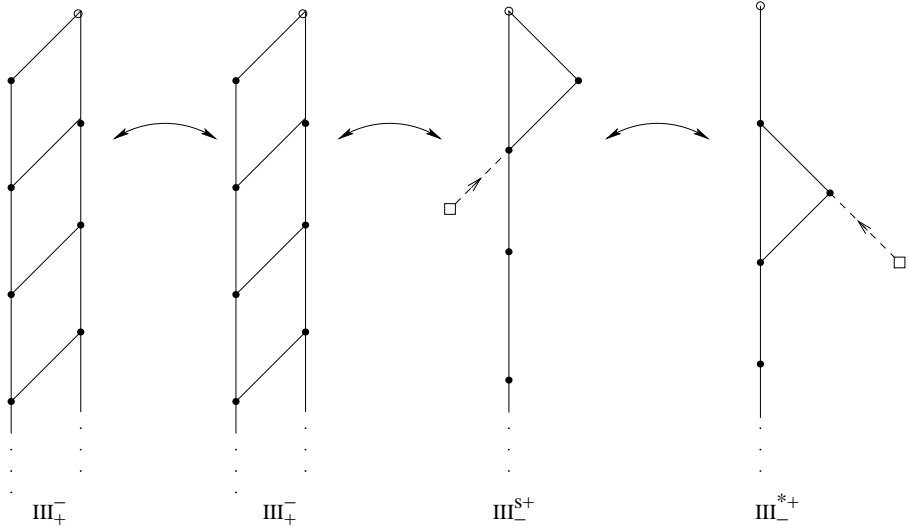}
  \caption{Spectral flow of type III diagrams from $h>0, |q|=2$ to
  $h<0, |q|=1$}
  \label{fig:star}
\end{figure}

We can now address the question of the orbit of a given
representation under spectral flow. 
Let us first assume that $h >0, q>1 $ and $\tfrac{h}{q} = k + \tfrac{1}{2}\in \mN +
\tfrac{1}{2}$. Gathering together the various pieces of information we see
that for $-n := \eta < 0$ the highest weight vector with respect to the
transformed modes has weight and charge
$(h + n q, q)$ and for  $n=\eta, k \geq n > 0$
it has weight and charge $(h - n q, q)$. For $\eta= k +1$ the values
of $(h^\eta,q^\eta)$ are $( h - \eta q + \tfrac{1}{2}, q
- 1) = ( - \tfrac{q -1}{2}, q-1)$. For $\eta = k +1 + n$ the new highest
weights and charges are given by $( - \tfrac{q -1}{2} -n(q-1), q-1)$.
Therefore what characterises the orbits is the charge $q$ of the
original highest weight vector. Spectral flow transformations of
representations with the highest weight vectors of the
form $\ket{q/2,q}$ create distinct orbits. Excluding the case
$|q|=1$ at the moment we can summarise the above by saying (remember $q>1$)
\begin{equation}
  \label{eq:orbit1}
  \ket{(\tfrac{q}{2})^\eta,\, q^\eta}_{\eta} = \begin{cases}
    \ket{-\tfrac{q-1}{2} - (\eta -1)(q-1),\, q-1}, & \text{for }
    \eta > 0 \\ 
    \ket{\tfrac{q}{2} - \eta q,\, q}, & \text{for } \eta < 0 .
  \end{cases}
\end{equation}
For the case $h>0, q<-1$, we obtain
\begin{equation}
  \ket{(-\tfrac{q}{2})^\eta,\, q^\eta}_{\eta} = \begin{cases}
    \ket{\tfrac{q+1}{2} - (\eta + 1)(q + 1),\, q +1}, & \text{for }
    \eta < 0 \\ 
    \ket{-\tfrac{q}{2} - \eta q,\, q}, & \text{for } \eta > 0.
  \end{cases}
\end{equation}
Thus the roles of $\eta>0$ and $\eta<0$ are interchanged. In figure
\ref{fig:typeIII} we 
have sketched which representations of this class are connected via
spectral flow. Actually there exists a special case when we start off
with a representation with $|q|=2$. These embedding diagrams are
mapped to the diagrams of type III$_-^{*\pm}$ under spectral flow if
we flow to negative values of $h$. This is depicted in
figure \ref{fig:star}. 

\begin{figure}[htbp]
  \centering
  \includegraphics[width=14cm]{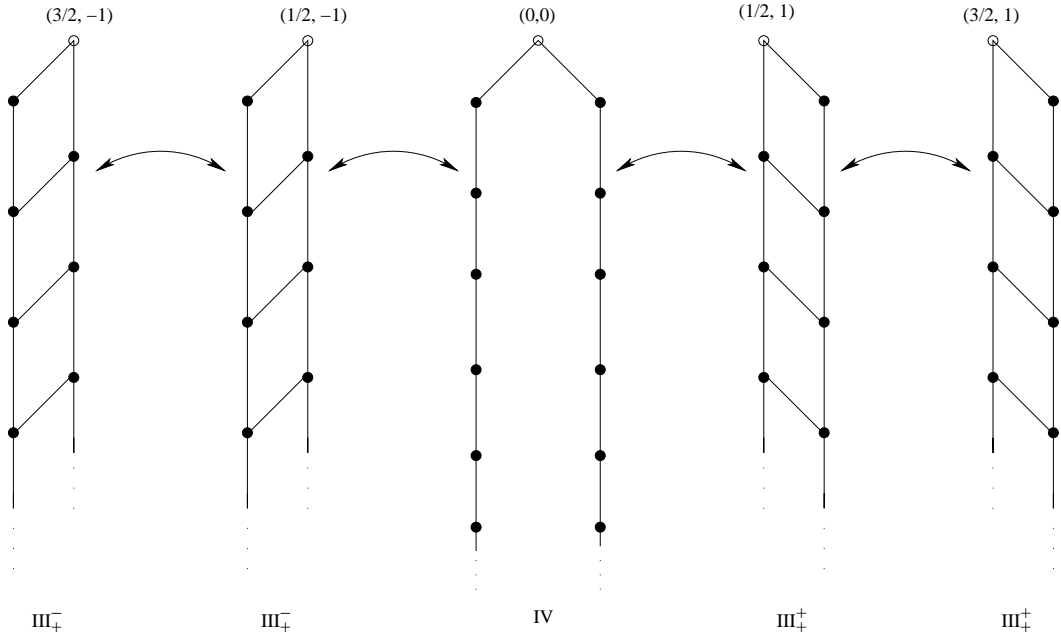}
  \caption{Spectral flow of the vacuum representation to
  representations with $h>0, |q|=1$}
  \label{fig:type0}
\end{figure}

The case $h=\tfrac{1}{2}$ and $q=1$ is special. For $\eta = 1$ it
is mapped to the vacuum representation: 
\begin{align}
  \alpha_1(L_0) & =(L_0-J_0 + \tfrac{1}{2})\ket{\tfrac{1}{2},1} = 0 \\
  \alpha_1(J_0) &= (J_0 -1)\ket{\tfrac{1}{2},1} = 0.
\end{align}
If we apply $\eta = 1$ on the vacuum representation, we obtain
the representation $h=\tfrac{1}{2}, q=-1$. This special case therefore
can be summarised as
\begin{equation}
\label{eq:null}
  \ket{0^\eta, 0^\eta}_{\eta}=
  \begin{cases}
    \ket{\tfrac{1}{2} + (\eta-1), -1} \text{ for } \eta >0 \\
    \ket{\tfrac{1}{2} + (1-\eta), 1} \text{ for } \eta <0. \\
  \end{cases}
\end{equation}
This transition under spectral flow is shown in figure \ref{fig:type0}.

\subsection{Unitarity}
\label{sec:unitarity}

The previous constructions show that a large class of
representations for $c=3$ are mapped to representations with $h<0$
which are manifestly not unitary. However the unitary representations are mapped
onto each other as the condition that $h>0$ is not sufficient to
guarantee unitarity.

In \cite{kent} the conditions for the existence of unitary
representations of the $N=2$ algebra were given. For $c=3$ the
relevant condition is given by
\begin{equation}
  (\ct,h,q) \text{ such that } g_n^A = 0, g_{n+\sgn(n)}^A < 0,
  f^A_{1,2}\geq0 \text{ for some }
  n\in \mZ + \tfrac{1}{2}.
\end{equation}
which translates into the following conditions for $h$ and $q$
\begin{align}
  \label{eq:2}
  (h - nq) &= 0 \\
  (h - (n + \sgn(n))q) & < 0 \\
  (1 - q^2) & \geq 0.
\end{align}
This implies that only diagrams of type I (figure \ref{fig:others})
for $|q|<1$ and diagrams of type III$_+^{\pm}$ with $|q|=1$ and the
vacuum representation (type IV) are unitary.

Under spectral flow unitary representations are mapped onto
each other. For the type IV and type III diagrams we can see this
behaviour in the way the spectral flow only maps the diagrams with
charge 0, 1, and $-1$ onto each other. 
In diagrams of type I we can see that unitarity is preserved by the
spectral flow from the following argument. 

Let us consider some representation $(h,q)$ such that
$\tfrac{h}{q} \in \mN + \tfrac{1}{2}$, $h>0$ and $0<q<1$. After spectral flow
towards 
lower values of $h$ the ratio of $h^\eta$ and $q$ will eventually be
given by $\tfrac{h^\eta}{q}=\tfrac{1}{2}$. At this point the
existing singular vector at level $\tfrac{1}{2}$ alters the
prescription of how the weight of the highest weight vector changes
and we obtain (c.f.\ equation \eqref{eq:22}) 
\begin{equation}
  \label{eq:52}
  \ket{h^\eta, q^\eta} = \ket{\tfrac{1-q}{2}, q-1}
\end{equation}
which defines again a unitary representation (note that $0<q<1$). For
$-1<q<0$ we obtain 
\begin{equation}
  \ket{h^\eta, q^\eta} = \ket{\tfrac{1+q}{2}, 1+q}.
\end{equation}

\section{Singular vectors under spectral flow}
\label{sec:singular-flow}

After writing down the orbits of representations it is important to
ensure that the singular vectors of a given representation transform
in a way consistent with the transformation properties of the highest
weight vector. 
The basic idea behind this section is the observation that the
spectral flow does not change the scalar product. 
In particular this implies that null vectors remain null vectors under
spectral flow. We will first discuss the situation for generic
singular vectors and then turn our attention to the representations
where subsingular vectors arise from spectral flow.

\subsection{Generic singular vectors}
\label{sec:sing-vect}

Let us consider a singular vector $\sN$ of some representation. If we
perform a spectral flow transformation on $\sN$, it will in
general not be mapped to a singular vector. However $\sN$ must be
mapped to a null vector as the inner product is invariant under
spectral flow. As only the  mode numbers of fermionic operators are
changed under spectral flow only some fermionic operator  can fail to
annihilate $\sN$ after a spectral flow transformation. If we look at
the case for $\eta=1$ we 
observe, comparing with equations \eqref{example} and
\eqref{example2}, that $\tgp{1/2}$ fails to annihilate $\sN$ and
the new singular vector is given by
\begin{equation}
  \label{eq:3}
  \tgp{1/2}\sN.
\end{equation}
Therefore singular vectors transform generically in the same way as the
highest weight state and the structure of the embedding
diagrams is (generically) unchanged. 

There are a few situations where the above considerations fail to give
meaningful results. The first situation where this method fails is the
one described preceding equations \eqref{eq:9} and \eqref{eq:10}. That
is, if the highest weight vector does not change under spectral
flow. The second situation can arise in representations with $h<0$ where
the number of charged singular vectors changes. 

\subsection{Subsingular Vectors}
\label{sec:subsingular}

In this section we want to show how spectral flow transformations
dictate the existence and form of subsingular
vectors in the representations under consideration. Generally there
are two classes of subsingular vectors: those with relative charge one
and those with higher relative charge. As the situations where they
arise are quite different we study both cases in turn. We will first
consider the subsingular vectors of charge one. They can arise if we flow
from a representation with $h>0$ to a representation with $h<0$.

\subsubsection{Subsingular vectors of charge one}
\label{sec:subs-vect-charge}

In this subsection we want to give evidence that the spectral flow
transformations which are shown in figure \ref{fig:typeIII} are
correct. In order to do so we have to explain how the subsingular
vector arises under spectral flow. We will first explain the spectral
flow from diagrams with $h>0$ to diagrams with $h<0$ and then turn our
attention to the spectral flow from the diagram with $h<0$ without a
subsingular vector to the diagram with $h<0$ with a subsingular
vector. 

\begin{figure}[htbp]
  \centering
  \input{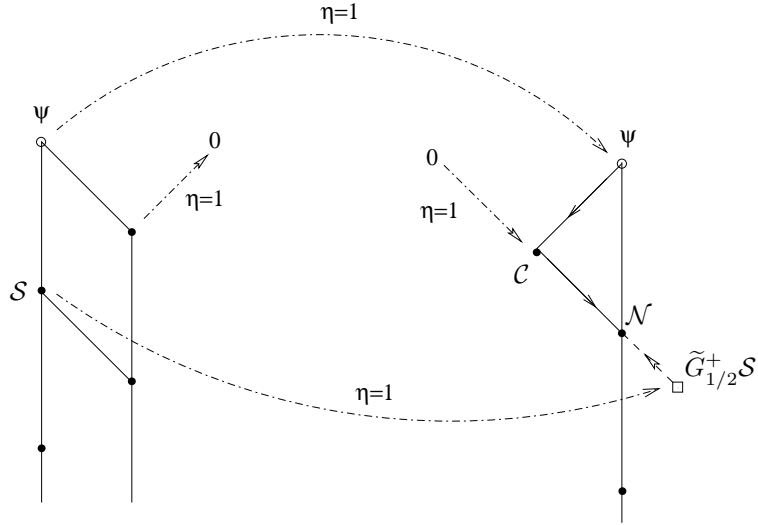}
  \caption{Spectral flow from singular to subsingular vectors}
  \label{fig:flow}
\end{figure}

Consider the spectral flow with flow parameter $\eta = 1$ of a
representation of type III$_+$ with $h=q/2$ and $q\geq2$ to a
representation of type III$_-$, see figure \ref{fig:flow}. In the
original representation the state $\Gp{-1/2}\ket{q/2,q}$ is a singular
vector 
and therefore, as explained in section \ref{sec:flow}, the highest
weight vector remains the highest weight vector after spectral flow. 
With respect to the transformed modes weight and charge are given by
\begin{align}
  \label{test}
  \wt{L}_{0}\ket{q/2,q} & = - \frac{q-1}{2}\ket{q/2,q}, \\
  \wt{J}_0\ket{q/2,q} & = (q - 1)\ket{q/2,q}.
\end{align}
Therefore the embedding diagrams change from an infinite number of
positively charged singular vectors to a finite number of negatively
charged singular vectors as is shown in diagram \ref{fig:typeIII}. 

Consider now the action of the spectral flow on the singular vectors.
The charged singular vectors are all descendants of the
vector $\Gp{-1/2}\ket{q/2,q}$. This vector is mapped to
$\tgp{1/2}\ket{q/2,q}$ and therefore is set to zero by the highest weight
condition. This implies that all its descendants are mapped to zero as
well. This is not true for the uncharged singular vector
at level $q$. If we call this vector $\sS$, by general arguments of
chapter \ref{sec:sing-vect} we know that $\sS$ must be mapped to
a null vector and furthermore $\tgp{1/2}\sS$ must span some submodule.
If we look at the embedding diagram of the representation after
spectral flow we observe that all singular vectors 
are descendants of $\sC=\tgm{-1/2}\ket{q/2,q}$. The inverse spectral flow
transformation maps $\sC$ to $\Gm{1/2}\ket{q/2, q}$ which is set
identically to zero. As $\sC$ is identically zero after the
inverse spectral flow transformation it follows that $\tgp{1/2}\sS$ can not be
a descendant of $\sC$. Nevertheless $\tgp{1/2}\sS$ is a null vector.
The resolution is that $\tgp{1/2}\sS$ is a subsingular
vector. More specifically $\tgm{1/2}\tgp{1/2}\sS$ is
given by the uncharged singular vector $\sN$, as can be seen as
follows:
Suppose first of all $\tgm{1/2}\tgp{1/2}\sS=0$. Then $\tgp{1/2}\sS$ is
itself a 
singular vector of charge $+1$ which is not allowed for a module with
a negative ratio of $\tfrac{h}{q}$. Therefore $\tgm{1/2}\tgp{1/2}\sS$ must not
vanish. However, all lowering operators applied to
$\tgm{1/2}\tgp{1/2}\sS$ vanish and therefore
$\tgm{1/2}\tgp{1/2}\sS$ is proportional to the uncharged
singular vector $\sN$. 

On the other hand $\tgp{1/2}\sS$ can not be a descendant of $\sN$
because $\sN$ is a descendant of $\tgp{-1/2}\psi$ which is
identically zero after the inverse spectral flow
transformation. $\sS$ is a descendant of $\tgp{1/2}\sS$ and
therefore if $\tgp{1/2}\sS$ were a descendant of $\sN$ so were
$\sS$. This would imply that $\sS$ had to vanish under the inverse
spectral flow transformation. That $\sS$ is indeed a
descendant of $\tgp{1/2}\sS$ can be seen as
follows. We compute $\tgm{-1/2}\tgp{1/2}\sS = (2\wt{L}_0 + \wt{J}_0 -
\tgp{1/2}\tgm{-1/2})\sS$. The commutator in the untilded modes is given
by $(2L_0 - J_0)\sS = 2q\sS$ and $\tgp{1/2}\tgm{-1/2}\sS =
\Gp{-1/2}\Gm{1/2}\sS=0$. Therefore we conclude that
$\tgp{1/2}\sS$ is a subsingular vector.

\begin{figure}[htbp]
  \centering
  \input{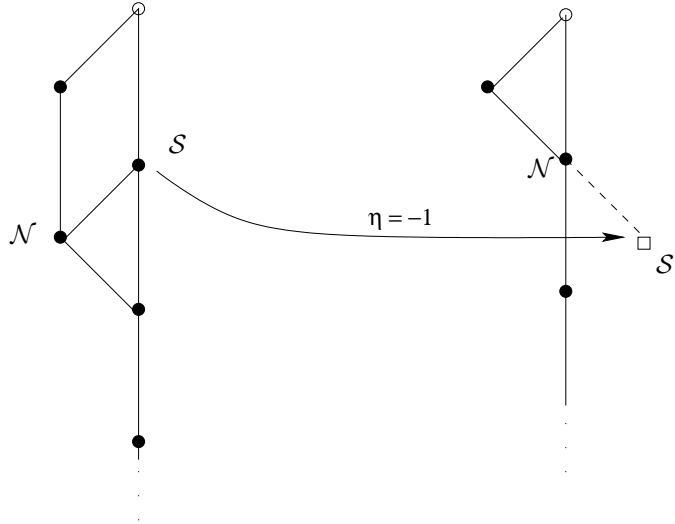}
  \caption{Spectral flow to a subsingular vector}
  \label{fig:flowm}
\end{figure}

In order to see how this matches with the spectral flow from
representations with $h<0$ let us start
with the representation $h=-\tfrac{3}{2}q$, $q \in \mN$.
As has been explained in section \ref{sec:reps} the number of charged
singular vectors varies for representations of type III$_-$ with
$h<0$. The 
representations with $h$ and $q$ as given above have two charged
singular vectors. The first one has level $|\tfrac{h}{q}| = 3/2$ and
the second one has level 
$|\tfrac{h}{q}| + (q-1) = \tfrac{1}{2} + q$. Let us call this charged
singular vector $\sN$. $\sN$ is a descendant of the first uncharged
singular vector $\sS$ as can be read off from figure
\ref{fig:flowm}. Comparing the levels of the singular 
vectors $\sN$ and $\sS$ we see that $\sN$ must be proportional to
$\Gm{-1/2}\sS$. Now we consider a spectral flow transformation with
flow parameter $\eta=-1$. Under this spectral flow transformation
$\sN$ is mapped to $\tgm{1/2}\sS$. Comparing this with the general
result of formula \eqref{eq:3} we observe that after spectral flow
$\tgm{1/2}\sS$ itself already defines an uncharged singular
vector. Therefore, under spectral flow $\sN$ is mapped to an uncharged
singular vector. The previous uncharged singular vector $\sS$ becomes
subsingular because
\begin{align}
  \tgp{1/2}\sS &= \Gp{3/2}\sS = 0\\
  \tgp{-1/2}\sS &= \Gp{1/2}\sS = 0 \label{eq:13}\\ 
  \tgm{1/2}\sS &= \Gm{-1/2}\sS = \sN.
\end{align}
$\sS$ is not a descendant of $\sN$ because $\tgp{-1/2}\sN =
\Gp{1/2}\sN =0$ although a lowering operator maps
$\sS$ to $\sN$. By general arguments $\sS$ is a null vector. Therefore it
has to be a subsingular vector since there exist no singular
vectors of which $\sS$ could be a descendant. 

Observe that the construction of the subsingular vector above depends
crucially on the fact that there exists an uncharged singular vector
whose $G^\pm_{-1/2}$ descendant is a charged singular vector.
In any representation with $h= -\frac{2l+1}{2}|q|$, $q\in \mZ$,
$l\in\mN$ there exists an uncharged singular vector which satisfies
this condition. This singular vector is mapped to an analogue of $\sS$
under spectral flow and enjoys the same properties, namely that the
application of $\wt{G}^\mp_{1/2}$ maps it to a singular vector,
however it is not the descendant of this singular vector. 
Unlike in the case described above the analogue of $\sS$ is for higher
values of $|h|$ a descendant  of some other singular vector of lower level. 
It is only for the case $l=1$ that there is no other singular
vector of which $\sS$ could be a descendant after spectral
flow. The only singular vector with lower level than $\sS$ is the
highest weight vector. 
Therefore only in this case the subsingular vector appears in the
embedding diagram because in all other cases it is 
a descendant state.

\subsubsection{Subsingular vectors of higher charge}
\label{sec:more-sub}

In diagrams of type III$_-^{*\pm}$ there exist subsingular vectors of higher
relative charge\footnote{For $c<3$ subsingular vectors of higher relative
charge were first discussed in \cite{rivera}. However the example
given there is not subsingular for $c=3$. The subsingular vector
constructed in \cite{rivera} is still a null vector for $c=3$ but
it becomes a descendant state of a singular vector.}.
The embedding diagrams
of the modules of type III$_-^{*\pm}$ are shown in 
figure \ref{fig:embeddingsub}.  The easiest example of this type of
subsingular vectors is given by  
\begin{equation}
  \label{eq:40}
  \sS = \Gm{-3/2}\Gm{-1/2}\ket{-3/2,1}.
\end{equation}
As one can check, $\sS$ is not a descendant of any singular vector, in
particular not a descendant of the first uncharged singular vector and
application of the operator $\Gp{1/2}$ maps $\sS$ to the charged
singular vector of the Verma module. 

Subsingular vectors of this type arise in all Verma modules with
$|q|=1$, $h=-\tfrac{2l + 1}{2}, l\geq1$. Before we comment on the
behaviour under spectral flow let us construct these subsingular
vectors for all representations in which they arise.

\noindent
Let us first define for convenience
\begin{equation}
  \label{eq:28}
 \Phi(-\tfrac{2l+1}{2},1):=
 \Gm{-\frac{2l + 1}{2}}\Gm{-\frac{2l - 1}{2}} \dots
 \Gm{-\frac{1}{2}}\ket{-\tfrac{2l+1}{2},1}. 
\end{equation}
We then observe 
\begin{align}
  \label{eq:29}
  \Gp{\frac{2l+1}{2}}\Phi(-\tfrac{2l+1}{2},1)  = 0 \\
  \{\Gp{\frac{2l-1}{2}},\Gm{-\frac{2l-1}{2}}\}\Phi(-\tfrac{2l+1}{2},1)
  = 0 \label{eq:30}.
\end{align}
Furthermore, inspection shows that 
\begin{equation}
  \label{eq:32}
  \Gp{\frac{2l -1}{2}}\Phi(-\tfrac{2l+1}{2},1) = \Gm{-\frac{2l -1}{2}} \dots
  \Gm{-\frac{1}{2}}\sN_1,  
\end{equation}
where $\sN_1$ denotes the uncharged singular vector at level one, see
figure \ref{fig:example}. These equations are proved in the
appendix. Therefore $\Phi(-\tfrac{2l+1}{2},1)$ defines a subsingular 
vector as it is a null vector and as it is not possible to reach
$\Phi(-\tfrac{2l+1}{2},1)$ from 
$\sN_1$, whereas applying an appropriate lowering operator maps
$\Phi(-\tfrac{2l+1}{2},1)$ 
to a descendant of $\sN_1$. 
However it is not possible to map
$\Phi(-\tfrac{2l+1}{2},1)$ to $\sN_1$ itself. 
In the example above
the descendant in question $\Gm{-1/2}\sN_1$ is the charged singular
vector of the Verma module. 
Observe that $\Phi(-\tfrac{2l+1}{2},1)$ is not annihilated by the
operators $\Gp{\frac{2l-3}{2}}, \dots, 
\Gp{\frac{1}{2}}$. Therefore, in general, it is not clear from the start
which vector actually defines the subsingular vector of lowest level.
The character formulae we calculate
in chapter \ref{sec:characters-c=3} seem to indicate that the lowest
lying subsingular vector is given by
\begin{equation}
  \label{eq:50}
  \sS_0:=\Gp{\frac{2l-3}{2}}\dots\Gp{\frac{1}{2}}\Phi(-\tfrac{2l+1}{2},1)
\end{equation}
and thus has always a relative charge of two.
In the examples we checked this vector was not a descendant of the
singular vector $\sN_1$, and therefore it was indeed the subsingular
vector of lowest charge and level.

\begin{figure}[htbp]
  \centering
  \input{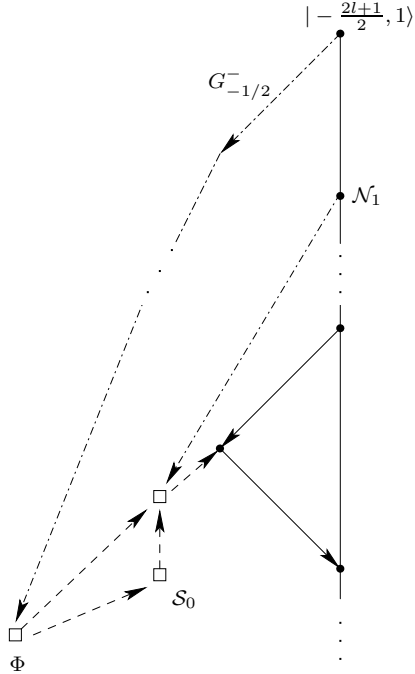}
  \caption{Connection between subsingular vectors for $|q|=1$}
  \label{fig:example}
\end{figure}

Vectors of the form as given by equation \eqref{eq:28} are mapped onto
each other under spectral flow. They are the images of the
singular vector $\Gm{-1/2}\ket{-1/2, 1}$ under spectral flow. As
mentioned in section \ref{sec:reps} there
exists only one charged singular vector in these embedding diagrams.
By the general construction described in section \ref{sec:subsingular}
this charged singular vector is mapped to an uncharged singular
vector under spectral 
flow with flow parameter $\eta=-1$. However a 
preimage of the ``new'' charged singular vector after spectral
flow must exist. This preimage is given by the subsingular vectors.
Therefore subsingular vectors of higher charge are needed for
consistency under spectral flow transformations.

An ambiguity in the embedding diagrams
arises when we consider subsingular vectors of higher charge.
All subsingular vectors of this kind are mapped to descendants
of the first uncharged singular vector $\sN_1$ however we can not
construct an operator which maps the subsingular vector to the
singular vector $\sN_1$ itself whereas the application of an
appropriate lowering operator maps the 
subsingular vector to the charged singular vector. We have
adopted the convention to connect the subsingular vector with the
singular vector of lowest level whose descendant can be reached by any
lowering operator.

\section{The Ramond algebra}
\label{sec:ramond-algebra}

In this section we will briefly comment on the spectral flow between the
Neveu-Schwarz and the Ramond (R) algebra. As has been remarked in 
e.g.\ \cite{schwimmer} the spectral flow connects the R and the NS algebra  and
they are essentially equivalent. 
The link between the highest weight representations is conveniently
easy as spectral flow with $\eta = \pm 1/2$ maps NS ground states to R
ground states. 

Consider some highest weight state $\ket{h,q}$. Under, say $\eta =
1/2$, the relevant modes are mapped to
\begin{align}
  \label{eq:6}
  \alpha_{1/2}(L_0)\ket{h,q} = (L_0 - \tfrac{q}{2} +
  \tfrac{1}{8})\ket{h,q} \\
  \label{eq:25}
  \alpha_{1/2}(J_0)\ket{h,q} = (J_0 - \tfrac{1}{2})\ket{h,q} \\
  \label{eq:26}
  \alpha_{1/2}(\Gp{1/2})\ket{h,q} = \Gp{0}\ket{h,q} \\
  \label{eq:27}
  \alpha_{1/2}(\Gm{1/2})\ket{h,q} = \Gm{1}\ket{h,q}
\end{align}
From this we see that the NS highest weight state remains a R highest
weight state, if we employ the condition that
$\Gp{0}\ket{h,q}=0$. This is one of the two possible choices of which
fermionic zero mode should annihilate the ground state. The other
choice has to be made for $\eta = -1/2$.
With this requirement the NS highest weight state is mapped to a R
highest weight state with eigenvalue $h - \tfrac{q}{2} + \tfrac{1}{8}$
and charge $q-\tfrac{1}{2}$. Therefore the embedding diagrams of the R
algebra have the same form as the embedding diagrams of the NS algebra.

\section{Characters for $c=3$}
\label{sec:characters-c=3}

As an additional consistency check we calculated the characters of the
NS algebra for some of the representations discussed previously and
determined their behaviour under spectral flow.

\subsection{The Characters}
\label{sec:characters}

The Characters for the $N=2$, $c=3$ algebra can be read off from the
embedding diagrams. We discuss here only the characters of the type III
diagrams and of the vacuum representation. To obtain the
remaining characters is a straightforward exercise.

A general character over a representation $\sV$ of the superconformal
algebra is defined as 
\begin{equation}
  \label{eq:15}
  \chi^{\vpp}_{\sV}(q,z):=\Tr{\sV} (q^{L_0-c/24}z^{J_0}),
\end{equation}
where $q:=e^{2\pi i \tau}$ and $z:=e^{2\pi i \nu}$.
The generic character of the Verma module $\sV_{h,Q}$ is given
by\footnote{In this section we will use a capital $Q$ to denote the charge.}
  (see for example \cite{Dobrev:1987hq, Kiritsis:1988rv,
    Dorrzapf:1997jh, Eholzer} where the characters for the minimal
  models are discussed.) 
\begin{equation}
  \label{eq:5}
  \chi^{\vpp}_{h,Q}(q,z) = q^{h-c/24} z^Q \prod_{n=1}^\infty \frac{
  (1+q^{n-\frac{1}{2}}z) (1+q^{n-\frac{1}{2}}z^{-1})}{(1-q^n)^2}.
\end{equation}
The character of the vacuum representation for $c=3$ is given by
\begin{equation}
  \label{eq:4}
  \chi^{\vpp}_{0,0}(q,z) = q^{-\frac{1}{8}}
  \prod_{n=1}^\infty \frac{
  (1+q^{n-\frac{1}{2}}z) (1+q^{n-\frac{1}{2}}z^{-1})}{(1-q^n)^2} 
\left(
  1- \frac{q^{\frac{1}{2}}z}{1+q^{\frac{1}{2}}z} -
  \frac{q^{\frac{1}{2}}z^{-1}}{1+q^{\frac{1}{2}}z^{-1}} \right)
\end{equation}
as we have to subtract the subrepresentations spanned by the singular
vectors from the generic character. As discussed in
\cite{Dorrzapf:1997jh, Eholzer} the character of a
submodule spanned by a 
charged singular vector of level $n$ is given by $q^n/(1+q^{n-n'})$
where $n'<n$ is 
the level of the uncharged singular vector which is connected to the
charged singular vector by some operator.
The characters for type III embedding diagrams without subsingular
vectors are given by
\begin{multline}
  \label{eq:7}
  \chi_{h,Q}^{III}(q,z) = q^{h-\frac{1}{8}}z^Q  \prod_{n=1}^\infty \frac{
    (1+q^{n-\frac{1}{2}}z) (1+q^{n-\frac{1}{2}}z^{-1})}{(1-q^n)^2} \\ 
  \times\left( 1 - q^{|Q|} -
    \frac{q^{\left|\frac{h}{Q}\right|}z^{\sgn(\frac{h}{Q})}}{ 1 + 
      q^{\left|\frac{h}{Q}\right|}z^{\sgn(\frac{h}{Q})}} + 
    \frac{q^{\left|\frac{h}{Q}\right|+|Q+\sgn(h)|}z^{\sgn(\frac{h}{Q})}}{
    1 +  
      q^{\left|\frac{h}{Q}\right|+|Q+\sgn(h)|-|Q|}
    z^{\sgn(\frac{h}{Q})}} \right). 
\end{multline}
We have to divide out the first charged and uncharged
singular vectors and add the second charged singular vector in
again. Otherwise we would subtract the charged singular vectors of
higher level
twice as they are descendants of both the first uncharged and charged
singular vectors, see figure \ref{fig:embeddingpositiv} and
\ref{fig:embeddingsub}. 

The characters of the embedding diagrams with subsingular vectors are
special. For the type III$^s$ diagrams the characters are given by 
\begin{multline}
  \label{eq:12}
  \chi_{h,Q}^{s}(q,z) = q^{h-\frac{1}{8}}z^Q  \prod_{n=1}^\infty \frac{
  (1+q^{n-\frac{1}{2}}z) (1+q^{n-\frac{1}{2}}z^{-1})}{(1-q^n)^2} \\ 
 \times\left( 1 - 
   \frac{q^{\frac{1}{2}}z^{\sgn(\frac{h}{Q})}}{ 1 + 
     q^{\frac{1}{2}}z^{\sgn(\frac{h}{Q})}} - \frac{q^{|Q| +
       \frac{1}{2}}z^{-\sgn(\frac{h}{Q})}}{1 +
     q^{\frac{1}{2}}z^{-\sgn(\frac{h}{Q})}} \right).
\end{multline}
The first fraction has the usual form and accounts
for the singular vectors which have to be divided out. 
The second fraction accounts for the subsingular vector which is
annihilated by the operator $G^{\pm}_{-1/2}$, as can be seen from
equation \eqref{eq:13}. Therefore we have to divide out the states
produced by the application of this particular operator. This
character can be seen as a generalisation of the character for the
vacuum representation as it has (sub)singular vectors of opposite
charge which have to be subtracted from the generic character. 

The characters for the embedding diagrams of type III$^*$ are more
involved. For $h=-\tfrac{2l+1}{2}$ and $Q=1$ they are given
by 
\begin{multline}
  \label{eq:59}
   \chi_{-\frac{2l+1}{2},1}^{*}(q,z) =
  zq^{-\frac{2l+1}{2}-\frac{1}{8}}  \prod_{n=1}^\infty \frac{ 
  (1+q^{n-\frac{1}{2}}z) (1+q^{n-\frac{1}{2}}z^{-1})}{(1-q^n)^2} \\ 
 \times\left( 1 - q - \frac{q^{2l} z^{-2}}{(1 + q^{\frac{2l+1}{2}}z^{-1}) (1
  + q^{\frac{2l-1}{2}}z^{-1})} + \frac{q^{2l+1} z^{-2}}{(1 +
  q^{\frac{2l+1}{2}}z^{-1}) (1 + q^{\frac{2l-1}{2}}z^{-1})}\right).
\end{multline}
All singular vectors are descendants of the first uncharged singular
vector and are thus subtracted by the first $q$ in \eqref{eq:59}. The
subsingular vector is annihilated by two charged operators 
as described in section \ref{sec:more-sub}. Therefore analogously to
the procedure for ordinary singular vectors 
both operators have to be divided out. In this case the modules created
by the singular and the subsingular vector have an overlap and in
order to prevent subtracting some vectors twice the correction
given by the last fraction has to be added in again.
In the easiest case for $h=-3/2$ and $Q=1$ the character formula reads
\begin{multline}
  \label{eq:14}
  \chi_{-3/2,1}^{*}(q,z) = zq^{-\frac{3}{2}-\frac{1}{8}}
  \prod_{n=1}^\infty \frac{ 
  (1+q^{n-\frac{1}{2}}z) (1+q^{n-\frac{1}{2}}z^{-1})}{(1-q^n)^2} \\ 
 \times\left( 1 - q - \frac{q^2 z^{-2}}{(1 + q^{\frac{1}{2}}z^{-1}) (1
  + q^{\frac{3}{2}}z^{-1})} + \frac{q^3 z^{-2}}{(1 + q^{\frac{1}{2}}z^{-1}) (1
  + q^{\frac{3}{2}}z^{-1})}\right)
\end{multline}
and the overlap is given by the
state 
\begin{equation}
  \label{eq:61}
  \Gm{-3/2}\Gm{-1/2}\sN_1 = 2(L_{-1} + J_{-1})\sS.
\end{equation}
For representations 
with higher values of $|h|$ this formula gets rather cumbersome but
can be calculated explicitly, most conveniently via the application of
spectral flow to equation \eqref{eq:61}.

\subsection{Spectral flow of characters}
\label{sec:char-flow}

As an additional consistency check we calculated the behaviour of the
characters under spectral flow transformations. 
If we consider equation \eqref{eq:15} the obvious way how characters
should  transform under spectral flow is given by
\begin{equation}
  \label{eq:20}
  \Tr{\sV_{h,Q}}(q^{\wt{L}_0-c/24}z^{\wt{J}_0}) =
  \Tr{\sV_{h^\eta,Q^\eta}}(q^{L_0-c/24}z^{J_0}).
\end{equation}
That is, the trace of the transformed operators over the original
representation should equal the character of the representation
defined by the eigenvalues $h^\eta$ and $Q^\eta$ of $\wt{L}_0$ and
$\wt{J}_0$, respectively as defined in  
equation \eqref{eq:23} and \eqref{eq:24}.
The term on the left hand side of equation \eqref{eq:20} can be
rewritten  to give (for $\eta = 1$)
\begin{equation}
  \label{eq:39}
  \Tr{\sV_{h,Q}}(e^{2 \pi i (L_0 - J_0 + \frac{1}{2} - \frac{1}{8})\tau} e^{2 \pi i (J_0 - 1)\nu})
  = \Tr{\sV_{h,Q}}( e^{2 \pi i(\tau/2 -\nu)} e^{2 \pi i (L_0 -\frac{1}{8})\tau}
  e^{2 \pi i J_0(\nu - \tau)}).
\end{equation}
If we define $\tilde{z}:=e^{2 \pi i(\nu - \tau)}=zq^{-1}$ the trace over the
shifted operators can be written as 
\begin{equation}
  \label{eq:44}
  \Tr{\sV_{h,Q}}(q^{\wt{L}_0-c/24}z^{\wt{J}_0})=
  \Tr{\sV_{h,Q}}(q^{L_0}\zt^{J_0}q^{-\frac{1}{2}}\zt^{-1}).
\end{equation}
The factor $q^{-\frac{1}{2}}\zt^{-1}$ gives just a constant
contribution to the trace.
If we substitute this expression into the generic character of equation
\eqref{eq:5} and then rewrite it again in terms of $q$ and $z$ we
obtain
\begin{align}
  \label{eq:62}
  q^{-\frac{1}{2}}\zt^{-1} \chi^{\vpp}_{h,Q}(q,\zt) &=
  q^{\frac{1}{2}}z^{-1} q^{h-Q-\frac{1}{8}} z^Q  \frac{1 + 
    q^{-\frac{1}{2}}z}{1+q^{\frac{1}{2}}z^{-1}} \prod_{n=1}^\infty \frac{
    (1+q^{n-\frac{1}{2}}z) (1+q^{n-\frac{1}{2}}z^{-1})}{(1-q^n)^2} \\ \nonumber
  & = q^{h-Q-\frac{1}{8}} z^Q \prod_{n=1}^\infty \frac{
    (1+q^{n-\frac{1}{2}}z) (1+q^{n-\frac{1}{2}}z^{-1})}{(1-q^n)^2} \\ \nonumber
  & = \chi^{\vpp}_{h^\eta, Q}(q,z)
\end{align}
where we have used the fact that $\tfrac{1 + q^{-\frac{1}{2}}z}
{1+q^{\frac{1}{2}}z^{-1}}=q^{-\frac{1}{2}}z$.

Now let us consider the spectral flow of the vacuum representation. To
be more explicit we choose $\eta = 1$ and therefore we expect to
obtain the character of the representation $h=1/2$, $Q=-1$. Indeed, we
get 
\begin{multline}
  \label{eq:54}
  q^{-\frac{1}{2}}\zt^{-1}\chi^{\vpp}_{0,0}(q,\zt) =    
  q^{\frac{1}{2}} z^{-1}
  q^{-\frac{1}{8}} \frac{1 + q^{-\frac{1}{2}} z}{1+q^{\frac{1}{2}}z^{-1}}
  \prod_{n=1}^\infty
  \frac{(1+q^{n-\frac{1}{2}}z)(1+q^{n-\frac{1}{2}}z^{-1})}{(1-q^n)^2} \\ 
  \times \left( 1- \frac{q^{\frac{1}{2}-1}z}{1+q^{\frac{1}{2}-1}z} -
  \frac{q^{\frac{1}{2}+1}z^{-1}}{1+q^{\frac{1}{2}+1}z^{-1}} \right)
\end{multline}
which can be rewritten as 
\begin{multline}
  \label{eq:53}
  q^{\frac{-1}{2}}\zt^{-1}\chi^{\vpp}_{0,0}(q,\zt) =  
  q^{\frac{1}{2}-\frac{1}{8}}z^{-1}
  \prod_{n=1}^\infty \frac{
    (1+q^{n-\frac{1}{2}}z)(1+q^{n-\frac{1}{2}}z^{-1})}{(1-q^n)^2} \\
  \shoveright{  \left( \frac{1}{1 + q^{\frac{1}{2}}z^{-1}} - \frac{q(1 +
        q^{\frac{1}{2}}z^{-1})}{(1+ q^{\frac{1}{2}}z^{-1})(1+
        q^{\frac{3}{2}}z^{-1})} \right)} \\ = q^{\frac{1}{2}-\frac{1}{8}}z^{-1}
  \prod_{n=1}^\infty 
  \frac{(1+q^{n-\frac{1}{2}}z)(1+q^{n-\frac{1}{2}}z^{-1})}{(1-q^n)^2} \\
  \shoveright{ \left(1 - q - 
      \frac{q^{\frac{1}{2}}z^{-1}}{1+q^{\frac{1}{2}}z^{-1}} +
      \frac{q^{\frac{5}{2}}z^{-1}}{1+q^{\frac{3}{2}}z^{-1}} \right) }\\
  = \chi^{\vpp}_{1/2, -1}(q,z).
\end{multline}

In the same fashion one can show that the other characters transform
appropriately under spectral flow. In particular we can show that
\begin{equation}
  \label{eq:11}
  q^{-\frac{1}{2}}\zt^{-1}\chi^{III}_{Q/2,Q}(q,\zt)
=\chi^{s}_{-(Q-1)/2,Q-1}(q,z)
\end{equation}
and
\begin{equation}
  \label{eq:21}
  q^{-\frac{1}{2}}\zt^{-1}\chi^{s}_{-1/2,1}(q,\zt)
=\chi^{*}_{-3/2,1}(q,z).
\end{equation}
As the calculations are essentially the same we will not demonstrate
them here. 
We therefore conclude that the spectral flow transforms the characters
among each other as expected.

\section{Unitary minimal models}
\label{sec:minimal}

\subsection{Definitions and embedding diagrams}
\label{sec:unitary-def}

As an example how this technique can be applied to other values of the
central charge, let us briefly demonstrate how spectral flow
transformations act on the embedding diagrams of the unitary minimal
models. The representation 
theory of the minimal models was extensively discussed, see for
example \cite{Feigin:1998sw,Semikhatov:1998gv,Semikhatov:1997pf,Dorrzapf:1997jh} or \cite{Kiritsis:1988rv,Dobrev:1987hq} for
earlier results. 

\begin{figure}[htbp]
   \begin{center}
     \includegraphics{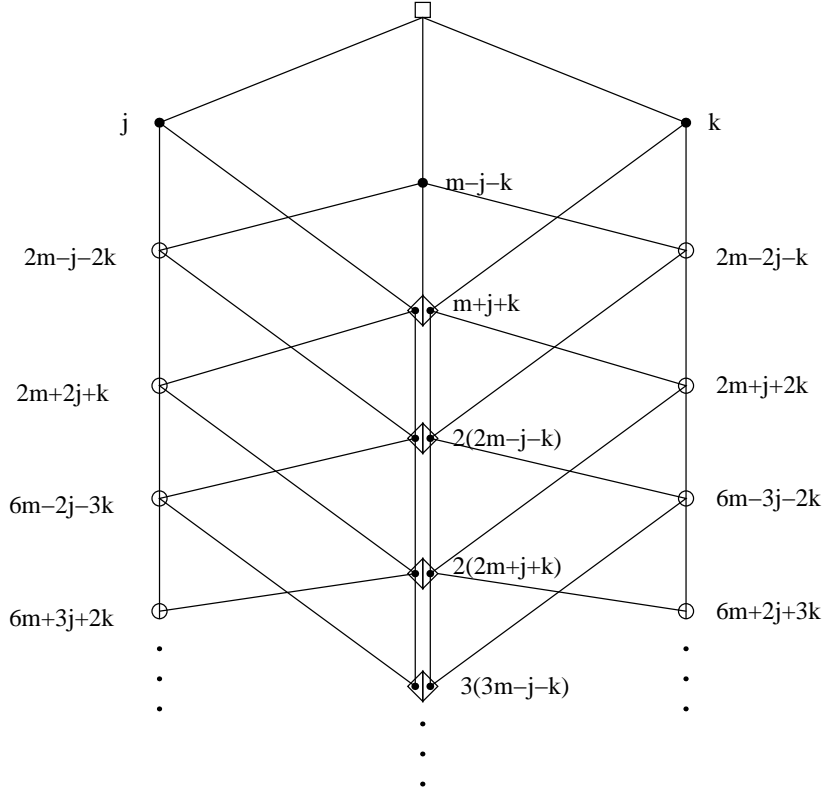}
     \caption{Embedding Diagram for the Unitary series with weights
       for the singular vectors.}
     \label{fig:unitary}
   \end{center}
 \end{figure}

Our definitions will closely follow \cite{Dorrzapf:1997jh}. The
unitary minimal models of the $N=2$ algebra can be parametrised by
 three parameters $m,j,k$ such that
 \begin{align}
   \label{eq:u1}
   c &=3(1-\tfrac{2}{m}), \qquad m\in \mN, m\geq 2 \\
   h &=\frac{jk-\frac{1}{4}}{m}, \qquad j,k \in \mN + \tfrac{1}{2},
   0<j,k,j+k\leq m-1\\
   q &=\frac{j-k}{m}.
 \end{align}
 The embedding diagrams for the minimal models were given in
 \cite{Dorrzapf:1997jh}. They have the form shown in figure
 \ref{fig:unitary}. In this diagram we denote the highest weight
 vector by a square, singular vectors corresponding to Kac-determinant
 vanishings are denoted by a filled circle and descendant singular
 vectors which do not correspond to Kac-determinant vanishings are
 denoted by an unfilled circle. 

The unitary models have the feature that they possess `uncharged
fermionic singular vectors' \cite{Dorrzapf:1997jh}. These (uncharged
singular) vectors possess charged singular vector descendants of only
positive or negative charge. In the embedding diagram the uncharged
fermionic singular vectors are
denoted by arrows pointing to the right or to the left, respectively.

\subsection{Spectral flow }
\label{sec:minimal-flow}

The action of the spectral flow on the highest weight states of
minimal models has been discussed in
\cite{Gaberdiel:1997kf}. Analogously to the $c=3$ case one has to
distinguish between a generic case where the highest weight vector
after spectral flow is given by a descendant of the highest weight
vector before spectral flow, and a special case where the descendant
which would be the new highest weight vector is itself a singular
vector. 
More precisely, a representation $(j,k)$ for fixed $m$ is generically
transformed under 
spectral flow with flow parameter $\eta=1$ to the representation
$(j+1, k - 1)$ where the highest weight state is then given by 
$\tgp{1/2}\ket{(j,k)}=\Gp{-1/2}\ket{(j,k)}$. 
If $k=\tfrac{1}{2}$, $\Gp{-1/2}\ket{(j,\tfrac{1}{2})}$ is a singular
vector and the state $\ket{(j,\tfrac{1}{2})}$ 
is mapped to the highest weight state after spectral flow. In this
case we find 
that the the representation $(j,\tfrac{1}{2})$ is mapped to the representation
$\alpha_{1}(j,\tfrac{1}{2})=(\tfrac{1}{2}, m-j-1)$. The same
phenomenon occurs for 
$\eta = -1$ and $j=\tfrac{1}{2}$. This situation is analogous to the
case for $c=3$, if $\Gm{-1/2}\ket{h,q}$ is a singular vector (c.f.\ section \ref{sec:rep-flow}). 

Let us first analyse the generic case $j,k \neq \tfrac{1}{2}$.
For a generic spectral flow transformation with $\eta=1$, the highest
weight state after spectral flow is given by $\tgp{1/2}\ket{(j,k)}$.
In this case a singular vector $\sN$ of the representation $(j,k)$ is
mapped to the singular vector $\tgp{1/2}\sN$. The levels of uncharged
singular vectors are derived in \cite{Dorrzapf:1997jh}. They are given by
$$
\Delta^0 = n(nm + j + k), \qquad n\in \mZ\backslash\{0\}.
$$
The combination $(j + k)$ is invariant under the generic spectral
flow, therefore  
the levels of uncharged singular vectors remain unchanged. This should
be the case as the level of an uncharged singular vector operator
is not changed by spectral flow. The levels of charged singular
vectors are given by
\begin{align}
  \label{eq:u2}
  \Delta^{+} &= k + n((n+1)m + j + k) \qquad n\in
  \mZ\backslash\{-1,0\}\\
  \Delta^{-} &= j + n((n+1)m + j + k) \qquad n\in
  \mZ\backslash\{-1,0\}  \nonumber
\end{align}
As has been discussed in section \ref{sec:rep-flow}, the levels of
charged singular vectors are altered under spectral flow according to 
$$
\alpha_{\eta}(L_0)\sN_\eta^{\pm} = (h_{\sN} \mp \eta)\sN_\eta^{\pm},
$$
where $\sN_\eta^{\pm}$ is the singular vector analogue of \eqref{eq:newhwv}. 
As we can read off from \eqref{eq:u2} the levels of the charged
singular vectors in the embedding diagrams transform accordingly. 
It can be shown that in the generic case, the spectral flow respects
the splitting of the uncharged singular vectors into left and right
fermionic uncharged singular vectors. The existence of uncharged
fermionic singular vectors is due to the fact, that some products
of singular vector operators vanish identically
\cite{Dorrzapf:1997jh}. Under (generic) spectral flow transformations
these vanishings are preserved. Therefore the splitting of the
uncharged singular vectors in left and right uncharged fermionic
vectors is preserved under generic spectral flow transformations.

Let us now analyse the spectral flow of $(j, \tfrac{1}{2})$ to
$(\tfrac{1}{2}, m-j-1)$. In the representation $(j, \tfrac{1}{2})$
the vector $\Gp{-1/2}\ket{(j, \tfrac{1}{2})}$ is singular and
therefore the highest weight vector remains a highest weight state
after spectral flow. 

In order to obtain a candidate for a singular vector from a singular vector
after spectral flow, we have to correct the shifted modes, and the new
singular vector would be given by $\tgp{1/2}\sN$. If we compare the
levels of singular vectors before and after spectral flow and keep in
mind that $\Gp{-1/2}\ket{(j, \tfrac{1}{2})}$ is mapped to
$\tgp{1/2}\ket{(j, \tfrac{1}{2})}$ and is therefore the new highest
weight condition, the images of the uncharged singular vectors are at
the levels of the positively charged singular vectors after spectral
flow and the images
of the negatively charged singular vectors are at the levels of the
uncharged singular vectors. The descendants of $\Gp{-1/2}\ket{(j,
\tfrac{1}{2})}$ vanish after spectral flow.
However the highest weight condition $\Gm{1/2}\ket{(j,
  \tfrac{1}{2})}$ is mapped to the singular vector $\tgm{-1/2}\ket{(j,
  \tfrac{1}{2})}$. Due to this fact the vectors $\tgp{1/2}\sN$ which
used to be singular vectors in the generic case are now mapped to
null vectors, which are {\em at least subsingular vectors}. I.e.\ in
general they are mapped to vectors of the appropriate level and charge
to be singular vectors, however their image under spectral flow in
general can be written as $\sN_{k}^{\rm{new}} = +\Theta_{-k}\ket{(1/2,
  m-j-1)} +  \Theta_{-k+1/2}\Gm{-1/2}\ket{(1/2, m-j-1)}$, where
$\Theta_{-k}$ is a (sub)singular vector operator. Therefore these
vectors are at least singular after dividing out $\Gm{-1/2}\ket{(1/2,
  m-j-1)}$. However the appearance of additional subsingular vectors
would result in additional submodules of all representations.
Therefore we can conclude that singular vector are flown to linear
combinations of (descendants of) singular vectors. 

As a final remark let us mention that we checked the transformation
properties of the characters of the minimal models (for their form
see e.g.\ \cite{Dorrzapf:1997jh}) under spectral
flow. Similarly to the $c=3$ case it can be shown that they transform
accordingly under spectral flow giving another argument that the
spectral flow respects the embedding structure.

\section{Conclusions}
\label{sec:conclusions}

We have analysed in detail the behaviour of representations and
singular vectors 
of the $N=2$ superconformal algebra with central charge $c=3$ under
spectral flow. The representations are mapped to representations of
the same kind in orbits determined by the charge of the highest weight
state. A careful analysis of the
spectral flow of singular vectors predicts subsingular vectors in
representations with $h<0$. It would be interesting to understand if
non-unitarity is a necessary condition for the existence of
subsingular vectors in the $N=2$ algebra.

Furthermore it was possible to construct subsingular vectors with
relative charge greater than two which were previously unknown. 
As an example how our technique can be used to analyse representations
with other values of $c$ we applied it to the unitary minimal models
and found the results consistent with the embedding diagrams.

As an additional check we determined the action of the spectral flow on
the characters and found complete agreement with the behaviour of the
embedding diagrams. To this end we had to construct the characters for
representations with subsingular vectors which, to our knowledge, have
not been written down in the literature before. 

As mentioned in the beginning we were not able to prove the actual
form of the embedding diagrams. However the behaviour under spectral
flow at least suggests that they are of the suggested form. 
As an additional check we performed some numerical tests.
We calculated the inner product matrices up to level four and compared
the dimension of their null-space with the dimension given by the null
vectors we assume to exist. The dimensions of the null-spaces matched
nicely with our expectations based on the embedding diagrams. We
performed this test on various representations up to level four and
found complete agreement.

\section*{Acknowledgements}
\label{sec:acknoledgements}

It is a pleasure to thank my PhD supervisor Matthias Gaberdiel for support
and helpful discussions, furthermore I would
like to thank 
Andreas Recknagel, Gerard Watts and in particular Matthias
D{\"o}rrzapf and Kevin Graham for 
interesting discussions and Matthew Hartley for a very helpful Perl
script. The computation of the inner product matrices were done using
Maple and \textsc{Reduce}.
This work was supported by a scholarship of the Marianne und Dr. Fritz
Walter Fischer-Stiftung and a Promotionsstipendium of the DAAD.

\begin{appendix}

\section{Proofs for section \ref{sec:more-sub}}
\label{sec:proofs}

We want to prove equations \eqref{eq:29} -- \eqref{eq:32}. 
Let $\Phi$ be defined as
\begin{equation}
  \label{eq:33}
   \Phi(-\tfrac{2l+1}{2},1):=
 \Gm{-\frac{2l + 1}{2}}\Gm{-\frac{2l - 1}{2}} \dots
 \Gm{-\frac{1}{2}}\ket{-\tfrac{2l+1}{2},1}
\end{equation}
as in equation \eqref{eq:28}. For ease of notation let us furthermore
define the abbreviation 
\begin{equation}
  \label{eq:37}
  \Gamma(2l-1) \ket{-\tfrac{2l+1}{2},1}:=  \Gm{-\frac{2l-1}{2}}
  \Gm{-\frac{2l - 3}{2}} \dots    
  \Gm{-\frac{1}{2}}\ket{-\tfrac{2l+1}{2},1}
\end{equation}
for the product of fermionic operators. This product of operators
satisfies:
\begin{equation}
  \label{eq:1}
  \Gp{\frac{2l+1}{2}} \Gamma(2l-1) \ket{h,q} = 0
\end{equation}
for any $h, q$, because the commutator of $\Gp{a}$ with any of the
operators $\Gm{-b}$ is 
proportional to some bosonic lowering operator $\Theta_{a-b}$. The 
commutator of this bosonic operator with any of the fermionic 
raising operators $\Gm{-c}$ is either proportional to some fermionic
lowering operator if $a>b+c$ which anticommutes with the other lowering
operators and annihilates the ground state or to some raising operator
$\Gm{a-b-c}$. As $a>0$ and because of the particular form of $\Gamma(2l-1)$,
$\Gm{a-b-c}$ will occur twice and therefore this expression will vanish.

The first equation we want to show is

\noindent
\textbf{Equation  \eqref{eq:30}}
  \begin{equation}
    \label{eq:41}
      \{\Gp{\frac{2l-1}{2}},\Gm{-\frac{2l-1}{2}}\}\Phi(-\tfrac{2l+1}{2},1)
  = 0.
  \end{equation}

\begin{proof}
  To show this we have to compute the anticommutator which is given
  by
  \begin{equation}
    \label{eq:17}
    \left(2L_0 + (2l-1)J_0 + \tfrac{(2l-1)^2 -1}{4} \right)
    \Phi(-\tfrac{2l+1}{2},1)
  \end{equation}
  The operators in the anticommutator evaluated on the given
  expressions are
  \begin{align}
    \label{eq:36}
    L_0  \Phi(-\tfrac{2l+1}{2},1)
    &= \left(\frac{(l+1)^2}{2} - \frac{2l + 1}{2}\right)
    \Phi(-\tfrac{2l+1}{2},1),\\
    J_0 \Phi(-\tfrac{2l+1}{2},1)   &= -l
    \Phi(-\tfrac{2l+1}{2},1). \label{eq:42}
  \end{align}
  Adding everything up proves equation \eqref{eq:30}.
\end{proof}
Now we turn our attention to

\noindent 
\textbf{Equation \eqref{eq:29}}
  \begin{equation}
    \label{eq:34}
    \Gp{\frac{2l + 1}{2}}\Phi(-\tfrac{2l+1}{2},1) = 0.
  \end{equation}  
\begin{proof}
We compute
  \begin{multline}
    \label{eq:35}
    \Gp{\frac{2l + 1}{2}}\Phi(-\tfrac{2l+1}{2},1) =  
    \Gp{\frac{2l + 1}{2}} \Gm{-\frac{2l + 1}{2}}
    \Gamma(2l-1)\ket{-\tfrac{2l+1}{2},1} = \\ 
    \left( \{\Gp{\frac{2l + 1}{2}}, \Gm{-\frac{2l + 1}{2}} \} - 
      \Gm{-\frac{2l + 1}{2}} \Gp{\frac{2l + 1}{2}} \right) 
   \Gamma(2l-1)\ket{-\tfrac{2l+1}{2},1} = \\ 
   \left(2L_0 + (2l+1)J_0 + \tfrac{(2l+1)^2 -1}{4}
    - \Gm{-\frac{2l + 1}{2}} \Gp{\frac{2l + 1}{2}} \right)
    \Gamma(2l-1)\ket{-\tfrac{2l+1}{2},1}.
\end{multline}
The last term of equation \eqref{eq:35} vanishes by equation \eqref{eq:1}. 
The terms stemming from the anticommutator vanish, as can be shown
using equations \eqref{eq:36} and \eqref{eq:42} and adjusting some
factors to take into account that the mode numbers differ slightly and
we can conclude
\begin{equation}
  \label{eq:38}
  \left(2L_0 + (2l+1)J_0 + \tfrac{(2l+1)^2 -1}{4} \right)
    \Gamma(2l-1)\ket{-\tfrac{2l+1}{2},1} = 0,
\end{equation}
which proves our statement.
\end{proof}
\noindent
Furthermore we have to prove equation \eqref{eq:32}, i.e.\ 

\noindent
\textbf{Equation  \eqref{eq:32}}
  \begin{equation}
    \label{eq:31}
    \Gp{\frac{2l -1}{2}}\Phi(-\tfrac{2l+1}{2},1) = \Gm{-\frac{2l -1}{2}} \dots
  \Gm{-\frac{1}{2}}\sN_1.
  \end{equation}

\begin{proof}
  The first step is to give the general form of the singular vector at
  level 1 of the Verma modules with $h=-\tfrac{2l + 1}{2}$ and $q=1$.
  Explicit calculations show that the first uncharged singular vector
  at level one is given by
  \begin{align}
    \label{eq:43}
    \sN_1 & = \left( 2L_{-1} - (2h + 1)J_{-1} -
      \Gm{-1/2}\Gp{-1/2}\right)\ket{h,1} \\
    & = \left(2L_{-1} +2l J_{-1}  -
      \Gm{-1/2}\Gp{-1/2}\right)\ket{-\frac{2l+1}{2},1}.
  \end{align}
  The next step is to evaluate 
  \begin{equation}
    \label{eq:46}
    \Gp{\frac{2l -1}{2}}\Phi(-\tfrac{2l+1}{2},1) 
  \end{equation}
  which gives 
  \begin{align}
    \label{eq:47}
    \Gp{\frac{2l -1}{2}}\Phi(-\tfrac{2l+1}{2},1)  &= (2L_{-1} + 2lJ_{-1})
    \Gm{-\frac{2l-1}{2}}\dots\Gm{-\frac{1}{2}}\ket{-\frac{2l+1}{2},1}
    \nonumber \\
    &- \Gm{-\frac{2l + 1}{2}} \left(2L_0 + (l-1)J_0 +
    \frac{(2l-1)^2-1}{4}\right) 
    \Gm{-\frac{2l-3}{2}}\dots\Gm{-\frac{1}{2}}\ket{-\frac{2l+1}{2},1} 
    \nonumber \\
    &+ \Gm{-\frac{2l+1}{2}}\Gm{-\frac{2l-1}{2}}\Gp{\frac{2l -1}{2}}
    \Gm{-\frac{2l-3}{2}}\dots\Gm{-\frac{1}{2}}\ket{-\frac{2l+1}{2},1}    
  \end{align}
 
The first line gives 
  \begin{equation}
    \label{eq:49}
    -2 \Gm{-\frac{2l + 1}{2}} \Gm{-\frac{2l-3}{2}} \dots
    \Gm{-\frac{1}{2}}\ket{-\frac{2l+1}{2},1} +\Gm{-\frac{2l-1}{2}} \dots
    \Gm{-\frac{1}{2}}\sN_1 
  \end{equation}
  because all other commutators of $L_{-1}$ and $J_{-1}$ with the
  fermionic operators are proportional to some other fermionic
  operator $\Gm{a-1}$. Apart from the first term this
  operator already exists in the row of operators and therefore the
  commutator vanishes. Only the first term and the last term are
  exceptional and remain.
  The second line of equation \eqref{eq:47} gives
  \begin{equation}
    \label{eq:48}
    2 \Gm{-\frac{2l + 1}{2}} \Gm{-\frac{2l-3}{2}} \dots
    \Gm{-\frac{1}{2}}\ket{-\frac{2l+1}{2},1}  
  \end{equation}
  The last line of equation \eqref{eq:47} is of the type
  $\Gp{\frac{2l+1}{2}}\Gamma(2l-1)\ket{h,q}$ and therefore vanishes. So
  we are left with 
  \begin{equation}
    \label{eq:16}
    (2 - 2) \Gm{-\frac{2l + 1}{2}} \Gm{-\frac{2l-3}{2}} \dots
    \Gm{-\frac{1}{2}}\ket{-\frac{2l+1}{2},1} + 
    \Gm{-\frac{2l-1}{2}} \dots \Gm{-\frac{1}{2}}\sN_1.
  \end{equation}
  The first two terms cancel out and thus we have proved our statement.
\end{proof}

\end{appendix}

\providecommand{\bysame}{\leavevmode\hbox to3em{\hrulefill}\thinspace}

\end{document}